\documentclass{jfm}
\usepackage[colorlinks=true,linkcolor=blue,citecolor=blue,urlcolor=blue]{hyperref}
\usepackage[doipre={doi:~}]{uri}
\usepackage{graphicx}
\usepackage{newtxtext}
\usepackage{newtxmath}
\usepackage{natbib}
\usepackage{hyperref}
\usepackage{amsmath}
\usepackage{xcolor}
\newcommand{\reddashline}{\textcolor{red}{\textbf{- - -}}}
\newcommand{\greendashdotline}{\textcolor{green}{\textbf{- $\cdot$ -}}}
\newcommand{\reddotline}{\textcolor{red}{\textbf{$\cdot$ $\cdot$ $\cdot$}}}

\hypersetup{
    colorlinks = true,
    urlcolor   = blue,
    citecolor  = blue,
}

\newcommand{\RomanNumeralCaps}[1]

\title{Close-contact melting on hydrophobic textured surfaces: Confinement and meniscus effects}

\author{Nan Hu\aff{1}\corresp{\email\href{nh0529@princeton.edu}{nh0529@princeton.edu}},
Li-Wu Fan\aff{2,3}\corresp{\email\href{liwufan@zju.edu.cn}{liwufan@zju.edu.cn}},
Xiang Gao\aff{2,4}
\and
Howard A. Stone\aff{1}\corresp{\email\href{hastone@princeton.edu}{hastone@princeton.edu}}
}

\affiliation{
\aff{1}Department of Mechanical and Aerospace Engineering, Princeton University, Princeton, NJ 08544, USA
\aff{2}State Key Laboratory of Clean Energy Utilization, Zhejiang University, Hangzhou 310027, People’s Republic of China
\aff{3}Institute of Thermal Science and Power Systems, School of Energy Engineering, Zhejiang University, Hangzhou, Zhejiang 310027, People’s Republic of China
\aff{4}Zhejiang Baima Lake Laboratory Co., Ltd., Hangzhou, Zhejiang 310051, People’s Republic of China
}

\begin{document}
\maketitle

\begin{abstract}
 We investigate the dynamics of close-contact melting (CCM) on ``gas-trapped" hydrophobic surfaces, with specific focus on the effects of geometrical confinement and the liquid-air meniscus below the liquid film. By employing dual-series and perturbation methods, we obtain numerical solutions for the effective slip lengths associated with velocity \(\lambda\) and temperature \(\lambda_t\) fields, across various values of aspect ratio \(\Lambda\) (defined as the ratio of the film thickness \(h\) to the structure's periodic length \(l\)) and gas-liquid fraction \(\phi\). Asymptotic solutions of \(\lambda\) and \(\lambda_t\) for \(\Lambda\ll 1\) and \(\Lambda \gg 1\) are derived and summarized for different surface structures, interface shapes and $\Lambda$, which reveal a different trend for $\lambda$ and $\Lambda \ll 1$ and the presence of a meniscus.
In the context of constant-pressure CCM, our results indicate that transverse-grooves  surfaces consistently reduced the heat transfer. However, longitudinal grooves can enhance heat transfer under the effects of confinement and meniscus when \(\Lambda \lessapprox 0.1\) and \(\phi < 1 - 0.5^{2/3} \approx 0.37\). For gravity-driven CCM, the parameters of \(l\) and \(\phi\) determine whether the melting rate is enhanced, reduced, or nearly unaffected. We construct a phase diagram based on the parameter matrix \((\log_{10} l, \phi)\) to delineate these three regimes. Lastly, we derived two asymptotic solutions for predicting the variation in time of the unmelted solid height.

\end{abstract}


\section{Introduction}
\label{sec:intro}

In the domain of melting, a phenomenon known as close-contact melting (CCM) stands out. Unlike conventional convection-driven melting with expanding molten liquid spaces, CCM involves a distinct process where the surface of an unmelted solid and a heating element are pressed together under external force. This results in the formation of a thin molten film flow between the solid and heating surfaces, facilitating solid-liquid phase change heat transfer at a low-Reynolds-number flow of the liquid. CCM can be observed in two primary modes: heat source-driven and unmelted solid-driven \citep{RN867}, depending on the external force applied to the object. The former mode refers to the scenario where the heated surface penetrates into an unmelted solid, which includes glacier drilling by heating \citep{RN369}, subtractive machining \citep{RN363}, and the `melt down' of nuclear fuel \citep{RN868}. The latter mode can be observed readily in daily life, such as when ice, butter, or chocolate melts on a heated substrate. Also, it is found in latent heat thermal energy storage \citep{RN390} and thermal management systems \citep{RN866}, where a considerable heat flux can be easily achieved. In the area of thermal energy storage, the melting rate is equivalent to the charging rate.

Given the widespread applications of CCM, it has been studied extensively for decades \citep{RN867}. These studies explore the effects of geometry \citep{RN377,RN392,RN444,RN427} and  boundary conditions for heating \citep{RN386,RN401,RN423,RN538,RN539,RN382,RN434}, with a primary focus on heat transfer characteristics. Recently, attention has shifted towards understanding the mechanisms of flow and heat transfer within the liquid film. Hu \textit{et al.} measured the thickness variation of the thin film for the first time, confirming a magnitude of $O(10^2)$ $\mu$m \citep{RN389}. Non-Newtonian features, including Bingham \citep{RN1038}, power-law \citep{RN1043}, and Carreau properties for the viscosity variations \citep{RN1042}, as well as temperature-dependent properties \citep{RN402}, have been considered to study their impact on film thickness variation and flow behavior. Strategies such as the use of permeable surfaces \citep{RN371}, slip surfaces \citep{RN373,RN1038}, and pressure-enhanced conditions \citep{RN866,RN1039} have been proposed to accelerate melting.

In addition, hydrophobic or superhydrophobic surfaces with microscale-trapped air can provide an effective slip length, on the order of $O(10^2)$ $\mu$m, for liquid transport \citep{Qur2008,RN1078}. This characteristic is considered a promising strategy for reducing liquid film thickness, thereby potentially enhancing heat transfer of CCM. Recently, Kozak theoretically studied a post-patterned hydrophobic surfaces and showed that  a  temperature slip length $\lambda^{*}_{t}$ (the definition is given in Sec XXX) greater than velocity slip length $\lambda^{*}$ \citep{RN1038}, leading to a reduced heat transfer; the calculation utilized the formulae $ \lambda^{*} / l^{*}=3 \sqrt{\pi/(1-\phi)}/16-3\ln (1+\sqrt{2})/(2 \pi) $ \citep{RN1097} and $\lambda^{*}=3\lambda^{*}_t/4$ \citep{RN1068}, where $^{*}$ represents dimensional quantities, $l^{*}$ is the period of the post pattern and $\phi$ is the liquid-gas area fraction. Although this 
work established a framework for analyzing velocity and temperature slip effects on CCM, the expressions for $\lambda^*$ relies on three assumptions: (1) shear flow in the liquid  above the surface, (2) a flat gas-liquid interface, and (3) perfect slip on the gas-liquid interface. Assumption (3) is satisfied  when $ h_s^{*} {\mu_l^{*}}/{\mu_g^{*}} \gg l^{*}$ \citep{RN1116,RN1045}, where $h_s^{*}$, $ {\mu_l^{*}}$, and ${\mu_g^{*}}$ denote micro-structure height, liquid viscosity, and gas viscosity, respectively. However, assumptions (1) and (2) may not apply to CCM due to the features of Poiseuille flow and deformation of the meniscus, leading to a complicated effective slip. The assumption of a constant ratio $\lambda_t^*/ \lambda^*$ is valid for large aspect ratios $\Lambda \equiv h^*/l^*$ \citep{RN1068}, where $h^*$ is the liquid film thickness. Furthermore, the slip lengths are  governed by the geometrical patterns of the micro-structure, which should be  evaluated and compared.

In the literature on slip, it has been shown that the velocity slip length $\lambda^{*}$ is influenced by confinement effects  \citep{RN1091,RN1096,RN1046,RN1047,RN1128,RN1048}. The asymptotic limit of slip length for flow past transverse slip regions inside a tube (radius $R^*$) has been determined  \citep{RN1091}, specifically $\lambda^{*}_{\perp} / l^{*} \sim  \ln \left(\sec \left(\phi \pi/2\right)\right)/(2\pi)$ when $l^{*}/R^{*} \ll 1$, while $\lambda^{*}_{ \perp}/R^{*} \sim \phi/(4-4\phi)$ when $l^{*}/R^{*} \gg 1$; this result indicates that the  velocity slip length can depend on the tube radius $R^{*}$ instead of the pattern period $l^{*}$. \cite{RN1096} numerically obtained for longitudinal grooves a variation of velocity slip length $\lambda_{\parallel} = \lambda_{\parallel}^{*}/l^{*}$ for various aspect ratios $\Lambda = h^{*} / l^{*}$ , where longitudinal means grooves oriented parallel to the flow, demonstrating that $\lambda_{\parallel}$ remains constant when $\Lambda \geq 10$, but shows a scaling law  $\lambda_{\parallel} \sim \Lambda$ when $\Lambda < 10$. Similar confinement effects on the effective slip length can be found for either longitudinal or transverse grooves with symmetrical or asymmetrical boundary conditions \citep{RN1046,RN1128} or on interfacial velocity at the liquid-gas interface \citep{RN1048}. An interesting fact related to confinement effects is that longitudinal grooves provide the largest slip lengths while transverse grooves yield the smallest ones in the limit of  $\Lambda \ll 1$ \citep{RN1047}. This contrasts with predictions for $\Lambda \gg 1$, where arrays of posts have a larger slip length than grooves.
Given the special feature of CCM that film thickness $h^{*}$ (also the channel height) is variable and sensitive to conditions, we can consider examples from previous work, where $l^{*} \sim 50 ~\mu$m and $h^{*} \sim 50 ~\mu$m ($\phi=0.99$), as well as $l^{*} \sim 16 ~\mu$m and $h^{*} \sim 80 ~\mu$m ($\phi=0.84$). In this case,  confinement effects may be non-negligible due to $1 \leq \Lambda \leq 5$.

At the same time, another issue arises concerning the curved gas-liquid interface due to the wide variation of pressure exerted on the liquid film in CCM. It has been demonstrated that meniscus protrusion into cavities will either increase or decrease the slip length $\lambda^{*}$ depending on the aspect ratio  $\Lambda$ \citep{RN1096,RN1129,RN1135,RN1062,RN1131}. When $\Lambda \gg 1 $, both slip lengths $\lambda^{*}_{\perp}$ and $\lambda^{*}_{\parallel}$ decrease along with a larger meniscus angle $\theta$ that protrudes into grooves \citep{RN1135}.  However, for \( \Lambda \ll 1 \), the slip length $\lambda^{*}$ can be increased by a meniscus for symmetrical boundaries  \citep{RN1062, RN1096}. 
Besides the velocity slip length \(\lambda^{*}\), confinement and meniscus effects also play an important role in the temperature slip length \(\lambda^{*}_{t}\)  \citep{RN1068,RN1062,RN1063}. The primary conclusions drawn are that the confinement effect results in a larger temperature slip length, while meniscus influences tend to reduce it. However, for CCM studies, the latter has not been considered \citep{RN1038}.

Here, we focus on parallel-groove textured surfaces as a promising candidate for CCM, owing to the theoretical limit of  a significant slip length for a given gas-liquid contact ratio \citep{RN1047}; detailed calculations for transverse grooves are given in the appendix. Furthermore, by incorporating double-reentrant structures, these surfaces can readily maintain the Cassie state for a wide range of liquids \citep{Liu2014,Wilke2022}
In \S\ref{sec_2}, we introduce the theoretical framework for close-contact melting, considering the influences of the velocity slip length $\lambda$ and the temperature slip length $\lambda_t$. Utilizing dual-series and perturbation methods, we present approaches to obtain numerical results for $\lambda$ and $\lambda_t$ for any aspect ratio $\Lambda$ and gas-liquid fraction $\phi$, for either flat or curved (a meniscus) gas-liquid interfaces. In \S\ref{sec_3}, we analyze the influences of $\Lambda$ and a meniscus on the slip length and derive asymptotic solutions for $\lambda_{\parallel,f}$, $\lambda_{\perp,f}$, $\lambda_{\parallel,c}$, and $\lambda_t$ as $\Lambda \to 0$ and $\Lambda \to \infty$, where $\parallel$ and $\perp$ represent longitudinal and transverse grooves, respectively, and $f$ and $c$ denote flat interfaces and menisci. A modified asymptotic solution for $\lambda_{\parallel,c}$ is proposed for a wide range of $\Lambda \gtrapprox 0.2$. We then discuss the slip effects on constant-pressure CCM, revealing the critical conditions for surface structures, $\Lambda$, and $\phi$ to achieve a faster melting rate. Additionally, we analyze gravity-driven CCM under limiting conditions derived from the dimensionless period length $l$, and construct a $l$-$\phi$ phase diagram to illustrate slip influences.

\section {Physical model and theoretical framework}
\label{sec_2}
\begin{figure}
\captionsetup{justification=justified}
\centerline{\includegraphics[width=\textwidth]{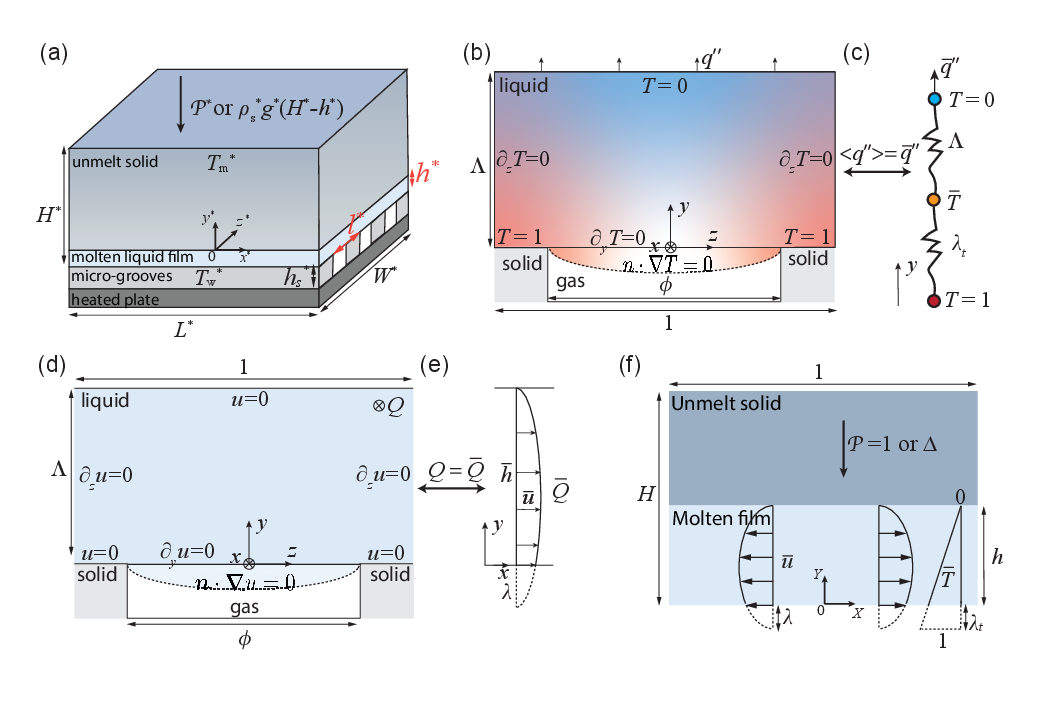}} 
	\caption{ (a) Close-contact melting of a cuboid-shaped unmelted solid with initial height $H_{0}^{*}$, length $L^{*}$ and width $W^{*}$ pressed downward by constant pressure $\mathcal{P}^{*}$ or self-weight $\rho_s^*g^*(H^*-h^*)$ on a heated microgrooved hydrophobic surface with characteristic period length $l^{*}$, where $h^*$ is the liquid film thickness; quantities with $^{*}$ are dimensional. Schematic diagrams of boundary conditions for 2D descriptions (b) or a 1D description (c) for the temperature distribution and temperature slip length $\lambda_{t}$; similarly, (d-e) characterize the velocity boundary conditions and velocity slip length $\lambda$ on the longitudinal grooves. It is noted that length is scaled by $l^{*}$  for convenience. (f) The dimensionless schematic diagram ($L^* \ll W^*$) by scaling as $X = x^*/L^*$ , $Y = y^*/h_0^*$ and $T = (T^*-T_m^*)/(T_w^*-T_m^*)$, showing the remaining solid height $H$ and film thickness $h$ is influenced by the effective velocity slip length $\lambda$ and temperature slip length $\lambda_{t}$. 
	} 
	\label{CCM_schematic}
\end{figure}

A typical CCM process on a micro-textured hydrophobic surface is sketched in Figure \ref{CCM_schematic}a. Here and in the following, the notation $^{*}$ represents a dimensional quantity. A cuboid-shaped solid phase change material (PCM) with length $L^{*}$, width $W^{*}$, and initial height $H_{0}^{*}$ is heated from below by a  textured (microgrooved) surface, characterized by a  wavelength $l^{*}$, at the wall temperature $T_{w}^{*}$. The solid PCM, initially at its melting point $T_{m}^{*}<T_{w}^{*}$, will continuously melt from the bottom under a certain pressure, which could be either a specific constant value $\mathcal{P}^{*}$ or the self-weight pressure of the PCM, $\rho_s^{*} g^{*} (H^{*}-h^{*})$, depending on the mode of CCM. 
Given the microgrooved structure (period length $l^*$ and gas fraction $\phi$), the mixed interface, composed of solid-liquid and gas-liquid contacts, when approximated with a one-dimensional description, { introduces an effective temperature slip $\lambda_{t}^{*}$ and velocity slip $\lambda^{*}$, which are described below and} significantly affect the fluid flow and heat transfer in the CCM process. 

\subsection{Scaling analysis and non-dimensionalisation}
 { Typically a thin melted film of liquid separates the substrate from the solid PCM. Hence, a lubrication-style analysis is suggested. The relevant physical parameters are the gravitational acceleration $g^{*}$, viscosity $\mu^{*}$, solid density $\rho_s^{*}$, liquid density $\rho_l^{*}$, liquid specific heat capacity $c_{p,l}^{*}$, thermal conductivity of liquid $k_l^{*}$, and latent heat of fusion $\mathcal{H}^*$. } By scaling pressure and stresses with $p_c^* = \rho_s^*g^*(H_0^*-h_0^*)$ for the gravity-driven mode or $p_c^* = \mathcal{P}^*$ for the constant-pressure mode, velocity with $u_c^* = {h_0^*}^2p_c^*/(\mu^*L^*)$, time  with {$t_c^* = L^*/u_c^* $}, and temperature with $T_w^{*}-T_m^{*}$, we can define dimensionless variables as 
\begin{equation}
    \begin{gathered}
    t =\frac{t^*}{t_c^*}, \quad X=\frac{x^*}{L^*}, \quad Y=\frac{y^*}{h_0^*}, \quad Z=\frac{z^*}{l^*}, \quad  u = \frac{u^*}{u_c^*}, \quad v = \frac{v^*L^*}{u_c^*h_0^*}, \quad w = \frac{w^*L^*}{u_c^*l^*}\\
    \quad P = \frac{P^*}{p_c^*}, \quad  T=\frac{T^*-T_m^*}{T_w^*-T_m^*},\quad h = \frac{h^*}{h_0^*}, \quad H =\frac{H^*}{H_0^*},  \quad \Delta = \frac{H^*-h^*}{H_0^*-h_0^*}, \\
     \mathscr{h} = \frac{h_0^*}{L^*}, \quad A = \frac{H_0^*}{L^*},  \quad \mathscr{l}=\frac{l^*}{L^*}, \quad l  =\frac{l^*}{h_0^*},\quad \Lambda=\frac{h^*}{l^*}=\frac{h}{l}
     \end{gathered}
     \label{nondimensional}
 \end{equation}
where $h_0^*$ is the initial film thickness identified below {as (\ref{h_0_star}) in \S \ref{average_formulation_ccm}} and aspect ratio $\Lambda$ is a key parameter related to confinement effect in the following discussion.
Hence, the dimensionless variables are denoted as the 
velocity vector $\boldsymbol{V}\equiv(u\boldsymbol{e}_X,v\boldsymbol{e}_Y,w\boldsymbol{e}_Z)$, temperature $T$,  pressure $P$, spatial gradient operator
$\boldsymbol{\nabla}\equiv \boldsymbol{e}_X\partial_X+\boldsymbol{e}_Y\partial_Y +\boldsymbol{e}_Z\partial_Z $, and  $\partial_t$ is the time derivative. The Reynolds number $Re$, Péclet number $Pe$, Prandtl number $Pr$, Eckert number $Ec$, Stefan number $Ste$, density ratio $\rho$, and hydrostatic pressure ratio $p_h$ are, respectively, defined as:
\begin{subequations}
    \begin{equation}
	Re \equiv \frac{\rho_l^*L^*u_c^*}{\mu^*}, \quad 
	Pe \equiv \frac{u_c^*L^*\rho_l^*c_{p,l}^*}{k_l^*}, \quad 
	Pr \equiv \frac{\mu^{*} c_{p,l}^{*}}{k_l^{*}}, 
\end{equation}
\begin{equation}
	Ec \equiv \frac{{u_c^*}^2}{ c_{p,l}^{*} \left(T_w^{*}-T_m^{*}\right)}, \quad 
	Ste \equiv \frac{c_{p,l}^{*} \left(T_w^{*}-T_m^{*}\right)}{\mathcal{H}^*}, \quad 
	\rho \equiv \frac{\rho_s^{*}}{\rho_l^{*}}, \quad p_h = \frac{\rho_l^*g^*h_0^*}{p_c^*}.
\end{equation}
\end{subequations}

In this case, the equations of continuity, momentum (in the $x$-, $y$-, and $z$-directions), and energy, including the dissipation of mechanical to thernmal energy in the bulk liquid and an energy balance at the melting front, are 
\begin{subequations}
\begin{equation}
 	\boldsymbol{\nabla} \cdot \boldsymbol{V}=0,
\end{equation}
\begin{equation}
	Re\mathscr{h}^2\partial_t u + Re\mathscr{h}^2 ({\boldsymbol{V}} \cdot \boldsymbol{\nabla}) {u} = -\partial_X P +  \mathscr{h}^2 \partial_X^2 u + \partial_Y^2 u + l^{-2} \partial_Z^2 u,
\end{equation}
\begin{equation}
	Re\mathscr{h}^4\partial_t v + Re\mathscr{h}^4 ({\boldsymbol{V}} \cdot \boldsymbol{\nabla}) v = -\partial_Y P +  \mathscr{h}^4 \partial_X^2 v + \mathscr{h}^2 \partial_Y^2 v + l^{-2} \mathscr{h}^2 \partial_Z^2 v + p_h,
\end{equation}
\begin{equation}
	Re\mathscr{h}^2\mathscr{l}^2 \partial_t w + Re\mathscr{h}^2 \mathscr{l}^2 ({\boldsymbol{V}} \cdot \boldsymbol{\nabla}) w = -\partial_Z P +  \mathscr{h}^2 \mathscr{l}^2 \partial_X^2 w + \mathscr{l}^2 \partial_Y^2 w + \mathscr{h}^2 \partial_Z^2 w ,
\end{equation}
\begin{equation}
\begin{split}
	Pe\mathscr{h}^2\partial_t T + Pe\mathscr{h}^2 ({\boldsymbol{V}} \cdot \boldsymbol{\nabla}) {T} 
	&= \mathscr{h}^2 \partial_X^2 T + \partial_Y^2 T + l^{-2} \partial_Z^2 T + Pr Ec \left[ 
	\left( \mathscr{h} \partial_X u \right)^2 + \left( \partial_Y u \right)^2 +
	\right. \\
	&\quad \left. \left( l^{-1} \partial_Z u \right)^2 + \left( \mathscr{h}^2 \partial_X v \right)^2 + \left( \mathscr{h} \partial_Y v \right)^2 +
	\left( l^{-1} \mathscr{h} \partial_Z v \right)^2 + 
	\right. \\
	&\quad \left. \left( \mathscr{h} \mathscr{l} \partial_X w \right)^2 + \left( \mathscr{l} \partial_Y w \right)^2 + \left( \mathscr{h} \partial_Z w \right)^2
	\right],
\end{split}
\end{equation}
\begin{equation}
	\frac{Ste}{\mathscr{h}(A-\mathscr{h}) Pe \rho} \partial_Y T(X,h) = \partial_t \Delta,
\end{equation}
\end{subequations}
where $\Delta$ is  the remaining height of the solid to be melted and  $\partial_t\Delta$ represents the melting rate. 

In accord with previous experimental and theoretical results \citep{RN386,RN1042}, the following well-validated assumptions are adopted: (a) the lubrication approximation is valid due to \( \mathscr{h}^2 \ll 1 \), (b) the flow is quasi-steady \(\partial/\partial{t}=0\) as a consequence of a low effective Reynolds number,  $Re\mathscr{h}^2 \ll 1$, (c) thermophysical properties are constant, (d) the solid-liquid interface is flat, i.e., \( h=h(t) \), (e) the hydrostatic pressure in the liquid film is neglected ($p_h \ll 1$), (f) heat convection is negligible $Pe\mathscr{h}^2 \ll 1$ when $Ste \leq 0.1$ for CCM \citep{RN1042,Ezra2024}, and (g) viscous dissipation is negligible due to \( Pr Ec \ll 1 \) for most PCMs and thermal conditions. Additionally we consider the geometrical conditions  \( \mathscr{l}^2 \ll 1 \) and $\mathscr{h} \ll A$. Consequently, {with $l  =l^*/h_0^*$ providing the corresponding transverse dimensions of the liquid flow channel,} the governing equations can be simplified considerably to:
\begin{subequations}
    \begin{equation}
	\partial_Xu+\partial_Yv=0
	\label{continuity}
\end{equation}
\begin{equation}
		0=	-\partial_XP+\partial_Y^2u+ l^{-2}\partial_Z^2u
		\label{momentum}
\end{equation}
\begin{equation}
		0=	\partial_YP
		\label{momentum-y}
\end{equation}
\begin{equation}
    0 = \partial_Z P
    \label{momentum-z}
\end{equation}
\begin{equation}
	0 = \partial_Y^2 T+l^{-2}\partial_Z^2T
	\label{temperature}
\end{equation}
\begin{equation}
	\partial_YT(X,h)= \frac{d\Delta}{d \tau}
	\label{interface},
\end{equation}
\end{subequations}
where  $\tau \equiv tSte/\left[\rho Pe\mathscr{h}(A-\mathscr{h})\right]$ is a re-scaled time, which recognizes that a balance of heat conduction and the heat of fusion control the dynamics.
Additionally, the force balance on the remaining solid is satisfied as
\begin{equation}
	\int_{-1/2}^{1/2} P dX = \Delta^\mathscr{c},
 \label{pressure}
\end{equation}
where $\mathscr{c} = 0$ represents the constant-pressure CCM mode and  $\mathscr{c} = 1$ represents a gravity-driven configuration. The goal to subsequent parts of Section 2 is determine the flow and temperature fields so as to find how the PCM height changes with time, i.e., $\Delta(\tau)$ or $H(\tau)$. A significant computational difficulty is to do this calculation for different groove configurations, i.e., gas-liquid fraction $\phi$ and dimensionless period length $l$ and dimensionless liquid film thickness $\Lambda$. The specific results are given in Section \S \ref{sec_3}.

\subsection{Average heat flux and effective thermal slip length}\label{thermal_slip_length}

In this section, a dual-series approach similar to  previous work \citep{RN1091,RN1096,RN1062} was adopted to give exact solutions for an effective thermal slip length $\lambda_{t}^{*}$ in the two-dimensional configuration. For convenience and consistency with previous work, the characteristic length $l^{*}$ {of the periodic topography of the substrate} is used for the length scale throughout this section, leading to dimensionless coordinates $(x,y,z)\equiv(x^*,y^*,z^*)/l^*$ and channel height $\Lambda \equiv h^*/l^*$. Consequently, the schematic diagram of the configuration is depicted in Figure \ref{CCM_schematic}b. The dimensionless variables $X$ and $Y$ introduced in the previous subsection will return when we obtain a 1D model of the problem in subsection \S \ref{average_formulation_ccm}.

In addition to the temperature distribution, our primary interest lies in the average heat flux $\left<q^{\prime\prime}\right> \equiv \int_{-1/2}^{1/2} -\partial_yT(\Lambda, z) dz$ through the upper boundary of the liquid film, as this parameter determines the melting rate as indicated by (\ref{interface}). Consequently, {the introduction of an effective thermal slip length $\lambda_t \equiv \lambda_t^*/l^*$ can reduce the two-dimensional problem $T(y,z)$ in the $yz$-plane to a one-dimensional problem $\bar T(y)$ in the $y$-direction, as illustrated in the schematic of the thermal transport in Figure \ref{CCM_schematic}c. In the following, $\lambda_{t,f}$ and $\lambda_{t,c}$ represent, respectively, the  thermal slip lengths of a flat and curved gas-liquid interface.

\subsubsection{ On the flat interface, $\lambda_{t,f}$}
\label{sec_lambda_t_f}

Based on (\ref{temperature}), the dimensionless form of the energy equation is rewritten as 
\begin{equation}
	\partial_{yy} T + \partial_{zz} T = 0.
\end{equation}
As depicted in Figure \ref{CCM_schematic}b, the boundary conditions are
\begin{subequations}
    \begin{equation}
	{T}(\Lambda,z)=0 
	\label{long_T_BC_1}
\end{equation} 
\begin{equation}
	\begin{cases}
		{T}(0,z)=1, & \phi /2 < |z| \le 1/2 \\ 
		\partial_y {T}(0,z)= 0, & |z| \le \phi /2
	\end{cases}
	\label{long_T_BC_2}
\end{equation} 
\begin{equation}
   \partial_z T(y,-1/2)= \partial_zT(y,1/2) = 0
    \label{T_partial_z},
\end{equation}
\end{subequations}where we assume that the heat flux vanishes at the gas-liquid boundary owing to the smallness of the ratio of the gas and liquid thermal conductivities. We can observe that the boundary conditions for this two-dimensional problem introduce the fraction of the boundary that involves the gas-liquid fraction $\phi$.  
Due to the linearity of the boundary-value problem, the temperature \(T\) can be divided into two components \(T_b\) and \(T_s\), namely $T(y,z)=T_b(y)+T_s(y,z)$, where \(T_b(y) = 1 - y/\Lambda\) for the homogeneous boundary condition, and \(T_s(y,z)=\tilde{T}(y,z)/\Lambda\) accounts for the hydrophobic surface. Hence, \(\tilde{T}\) also satisfies the Laplace equation as 
\begin{equation}
	\partial_{yy} \tilde{T} + \partial_{zz} \tilde{T} = 0.
\end{equation}
The boundary conditions for $\tilde{T}$ are
\begin{equation}
	\tilde{T}(\Lambda,z)=0 
	\label{long_T_BC_1_imhomo}
\end{equation} 
\begin{equation}
	\begin{cases}
		\tilde{T}(0,z)=0, & \phi /2 < |z| \le 1/2; \\ 
		-1 + \partial_y\tilde{T}(0,z) = 0, & |z| \le \phi /2.
	\end{cases}
	\label{long_T_BC_2_imhomo}
\end{equation} 
The general solution \(\tilde{T}\), accounting for condition  (\ref{T_partial_z}), is
\begin{equation}
	\begin{aligned}
		\tilde{T} (y, z) = c_0 \frac{y}{\Lambda} + d_0 + \sum_{n=1}^\infty \left[c_n \cosh \left(2\pi n y\right) + d_n \sinh \left(2\pi n y\right)\right] \cos \left(2\pi n z\right),
	\end{aligned}
 \label{T_general_form}
\end{equation}
where \(c_n\) and \(d_n\) for $n \in [0, \infty)$ are constants to be determined for given $\phi$ and $\Lambda$. Following the standard procedure of a dual-series method to numerically solve (\ref{T_general_form}) as detailed in Appendix \S\ref{Procedure_lambda_t_f}, $c_n$ and $d_n$ for $n \in [0, N-1]$ can be obtained with a numerical truncation for large enough $N$.
Therefore, the average heat flux $\left<q^{\prime\prime}\right>$ can be calculated by substituting $c_0$ as 
\begin{equation}
    \left<q^{\prime\prime}\right> = -\int_{-1/2}^{1/2} \partial_yT(\Lambda, z) dz = \frac{1}{\Lambda} - \frac{c_0}{\Lambda^2}.
\end{equation}
The definition of an effective thermal slip length $ \lambda_{t}$ representative of the composite flat liquid-gas surface and solid-liquid surface \citep{RN1068} is 
\begin{equation}
\lambda_{t} \equiv \frac{T_b(y=0)- \left<T(y=0)\right>}{\left<\partial_y T(y=0)\right>} = \frac{1-\int_{-1/2}^{1/2}(1+\tilde{T}(y=0)/\Lambda) dz}{\int_{-1/2}^{1/2} \partial_yT(0, z) dz}.
\end{equation}
With the results of $1-\int_{-1/2}^{1/2}(1+\tilde{T}(y=0)/\Lambda) dz= c_0/\Lambda$ due to $c_0 = -d_0$ from Appendix \S\ref{Procedure_lambda_t_f} and $\int_{-1/2}^{1/2} \partial_yT(0, z) dz= \left<q^{\prime\prime}\right>$, we find the effective thermal slip length $ \lambda_{t,f}$ for a flat, composite gas-liquid interface
\begin{equation}
	\begin{aligned}
		\lambda_{t,f} = \frac{{\Lambda}c_0}{\Lambda-c_0}.
	\end{aligned}
    \label{14}
\end{equation}

\subsubsection{On the curved interface, $\lambda_{t,c}$}\label{sec_lambda_t_c}
We next consider how the deformation of the gas-liquid interface affects the thermal slip length. A curved liquid-air interface is formed by the pressure difference between liquid \( P_l^{*} \) and gas \( P_g^{*} \), given as a curvature \( R^{*} = \sigma^{*}/(P_l^{*} - P_g^{*})\),
where \(\sigma^{*}\) is the surface tension. Hence, the meniscus  can be described by the geometric relationship 
\begin{equation}
y = \sqrt{R^2 - \phi^2 / 4} - \sqrt{R^2 - z^2},
\end{equation}
where \(R = R^{*} / l^{*}\). For \(R \gg \phi / 2\), it allows that
\begin{equation}
y = \frac{1}{8R} \left(4z^2 - \phi^2\right) + \mathcal{O}\left(\frac{1}{R^3}\right) = - \epsilon \eta(z) \quad \text{for} \quad |z| \leq \phi / 2,
\end{equation}
after defining \(\epsilon = 1 / (8R)\), which will be assumed $\ll 1$,  and \(\eta(z) = \phi^2 - 4z^2\). It should be noted that the protrusion angle \(\theta\) satisfies \(R \sin \theta = \phi / 2\), which leads to a maximum  \(\epsilon = \sin\theta/4\phi = 0.217\) for \(\phi = 0.2\) and \(\theta = 10^{\circ}\). Again, neglecting the influence of the conductivity of the gas phase, then we have the condition of no heat flux {across the meniscus}, and the temperature should satisfy
\begin{equation}
	\boldsymbol{n} \cdot \boldsymbol{\nabla}T = 0 \quad \text{at} \quad y =-\epsilon\eta(z),
\end{equation}
which is equivalent to 
\begin{equation}
	\partial_y T(-\epsilon\eta, z)+\epsilon \eta^{\prime} \partial_z T(-\epsilon\eta, z)=0,
	\label{meniscus no heat flux}
\end{equation}
where $\eta^{\prime} \equiv \mathrm{d} \eta/\mathrm{d}z =-8z$. Since $\epsilon \ll 1$, $T(-\epsilon\eta, z)$ can be obtained from $T(0, z)$ by Taylor expansion as $T(-\epsilon \eta, z)=T(0, z)-\epsilon \eta \partial_y T(0, z)+O\left(\epsilon^2\right)$, leading to
\begin{equation}
		\begin{aligned}
			& \partial_y T(-\epsilon\eta, z)=\partial_y T(0, z)-\epsilon \eta \partial_{yy} T(0, z)+O\left(\epsilon^2\right), \\
			& \partial_z T(-\epsilon\eta, z)=\partial_z T(0, z)+O(\epsilon).
			\label{meniscus_no_heatflux_boundary}
		\end{aligned}	
\end{equation}
Then, substitution of (\ref{meniscus_no_heatflux_boundary}) into (\ref{meniscus no heat flux})  yields
\begin{equation}
	\partial_y T(0, z)-\epsilon \eta \partial_{yy} T(0, z)+\epsilon \eta^{\prime} \partial_z T(0, z)+O\left(\epsilon^2\right)=0.
\end{equation}
Hence the boundary conditions are
\begin{equation}
\label{T_curved_BC}
	\begin{aligned}
		& T(\Lambda, z)=0 \\
		& \left\{\begin{array}{ll}
			T(0, z)=0, & \phi/2<|z| \leqslant 1 / 2; \\
			\partial_y T(0, z)-\epsilon \eta \partial_{y y} T(0, z)+\epsilon \eta^{\prime} \partial_z T(0, z)=0, & |z| \leqslant \phi/2.
		\end{array} \right.
	\end{aligned}
\end{equation}
After a perturbation analysis by substituting $T=T^{(0)}+\epsilon T^{(1)}+O\left(\epsilon^2\right)$ (details provided in Appendix \ref{Procedure_lambda_t_c}), we find that the presence of a curved gas-liquid interface has no influence on the average heat flux and thermal slip length, which means
\begin{equation}
\lambda_{t,f}=\lambda_{t,c}\equiv \lambda_t.
\end{equation}
We have now completed the necessary aspects of the thermal problem inside the thin film to estimate the time variations of the phase change material, which will be described below \S \ref{average_formulation_ccm}.

\subsection{Flow rate - pressure gradient relationship and 
velocity slip length}\label{slip_length}
We now consider the flow in the thin film since we need the pressure distribution to complete the force balance on the phase change material.
Based on (\ref{momentum})-(\ref{momentum-z}), the main liquid flow is along the $x$ direction as indicated in Figure \ref{CCM_schematic}d. Similar to \S \ref{thermal_slip_length}, in addition to obtaining the velocity field $u$, we also focus on the variation of the flow rate-pressure gradient relationship $Q-\partial_x P$ due to the presence of trapped gas, which will influence the pressure distribution within the liquid film and the liquid film thickness during the CCM process. By introducing the velocity slip length $\lambda \equiv \lambda^*/l^*$ to enable an equivalent flow rate $\bar Q = Q$ as demonstrated in Figure \ref{CCM_schematic}e, this allows a dimensional reduction from $u(y,z)$ to $\bar u(y)$. In the following, $\lambda_{\parallel,f}$ and  $\lambda_{\parallel,c}$ represents the velocity slip lengths by flat and curved gas-liquid interfaces, respectively, while the main liquid flow is parallel to the orientation of the grooves.  

\subsubsection{Velocity slip length on the flat interface, $\lambda_{\parallel,f}$}\label{sec_slip_length_f}
Consequently, the dimensionless form of the momentum equation is 
\begin{equation}
\partial_{yy}u+\partial_{zz}u=\partial_xP,
\end{equation}
where $u={u}^{*} \mu^{*} L^{*}/({{l}^{*}}^2p_c^{*})$.
Due to the linearity, the velocity $u(y,z)$ can be divided into two components $u_b(y)$ and $u_{s}(y,z)$, where $u_b(y) = -\partial_xP\left(\Lambda y-y^2\right)/2$ accounts for the flow without slip and $u_s(y,z)=-\partial_xP\Lambda\tilde{u}/2$ accounts for slip of the hydrophobic surface. Hence, substituting $u(y,z)= -\partial_xP\left(\Lambda y-y^2+\Lambda\tilde{u}(y,z)\right)/2$ leads to a boundary-value problem now with homogeneous boundary conditions as
\begin{equation}
	\partial_{yy} \tilde{u}+\partial_{zz} \tilde{u}=0.
\end{equation}
The corresponding boundary conditions are
\begin{equation}
	\tilde{u}(\Lambda, z)=0,
	\label{long_BC_1}
\end{equation} 
\begin{equation}
	\begin{cases}\tilde{u}(0, z)=0, & \phi /2 <|z| \le 1/2; \\ 1+\partial_{y} \tilde{u}(0, z)=0, & |z|\le \phi/2.\end{cases}
	\label{long_BC_2}
\end{equation} 
This problem is effectively identical to the thermal problem in subsection \S \ref{sec_lambda_t_f}. The general solution of $\tilde{u}(y,z)$ with periodic boundary conditions at $z = \pm 1/2$ is
\begin{equation}
	\begin{aligned}
		\tilde{u} (y,z) =  r_0 +q_{0}\frac{y}{\Lambda}+\sum_{n=1}^\infty \left[r_n\cosh \left(2\pi n y\right)+q_n \sinh \left(2\pi n y\right)\right] \cos \left(2\pi n z\right),
	\end{aligned}
\end{equation}
where constants $r_n$ and $q_n$ ($n \in [0, N-1]$), dependent on $\phi$ and $\Lambda$, also can be numerically determined similar to the dual-series approach in Appendix \S \ref{Procedure_lambda_perp_f}.
Consequently, the increased flow rate $Q_d$ due to the presence of slip can be obtained via
\begin{equation}
	Q_d = \int_{0}^{\Lambda}\int_{-1/2}^{1/2} u_s \mathrm{d}z\mathrm{d}y =\int_{0}^{\Lambda}\int_{-1/2}^{1/2} -\frac{dP}{dx}\frac{\Lambda\tilde{u}}{2} \mathrm{d}z\mathrm{d}y= -\frac{dP}{dx}\frac{{\Lambda}^2r_0}{4}.
\end{equation}
On the other hand, for a pressure-driven channel flow with one boundary of slip length $\lambda$, the increased flow rate $Q_d$ can be written as
\begin{equation}
	Q_d =  -\frac{dP}{dx}\frac{{\Lambda^3\lambda}}{4\Lambda+4\lambda}.
\end{equation}
Comparing the last two expressions, , the effective velocity slip length $\lambda_{\parallel,f}$ on the longitudinal grooves for a flat liquid-gas interface is
\begin{equation}
	\lambda_{\parallel,f} = \frac{\Lambda r_0}{\Lambda-r_0}.
    \label{30}
\end{equation}
A reader should notice that the detailed boundary value problems for the thermal and velocity fields, and corresponding dual integral equations solutions, show that the two problems are identical, $r_n=-c_n$ and $r_0=c_0$. Thus, $\lambda_{\parallel,f}=\lambda_{t,f}$ as shown from (\ref{30}) and  (\ref{14}). 
\subsubsection{Velocity slip length on the curved interface, $\lambda_{\parallel,c}$} 
\label{sec_slip_length_c}
To account for a curved gas-liquid interface, we consider the configuration  similar to \S \ref{sec_lambda_t_c}. 
 In order to satisfy the condition of a zero shear stress at the steady-state gas-liquid interface, the velocity should satisfy
\begin{equation}
	\boldsymbol{n} \cdot \boldsymbol{\nabla}u = 0 \quad \text{at} \quad y =-\epsilon\eta,
\end{equation}
which is equivalent to 
\begin{equation}
	\epsilon \eta^{\prime} \partial_z u(-\epsilon\eta, z)+\partial_y u(-\epsilon\eta, z)=0.
	\label{meniscus shear free}
\end{equation}
Since $\epsilon \ll 1$, $u(-\epsilon\eta, z)$ can be obtained from $u(y, z)$ by Taylor expansion as $u(-\epsilon\eta, z)=u(0, z)-\epsilon \eta \partial_y u(0, z)+O\left(\epsilon^2\right)$ as 
\begin{subequations}
    \begin{equation}
			 \partial_y u(-\epsilon\eta, z)=\partial_y u(0, z)-\epsilon \eta \partial_{yy} u(0, z)+O\left(\epsilon^2\right), \label{meniscus shear free boundary_y}
\end{equation}
\begin{equation}
    \partial_z u(-\epsilon\eta, z)=\partial_z u(0, z)+O(\epsilon).
    \label{meniscus shear free boundary_z}
\end{equation}
\end{subequations}
\\
Then, substitution of (\ref{meniscus shear free boundary_y}) and (\ref{meniscus shear free boundary_z}) into (\ref{meniscus shear free})  yields
\begin{equation}
	\partial_y u(0, z)-\epsilon \eta \partial_{yy} u(0, z)+\epsilon \eta^{\prime} \partial_z u(0, z)+O\left(\epsilon^2\right)=0.
\end{equation}
Hence, the boundary conditions are
\begin{equation}
\label{velocity_curve_BC}
	\begin{aligned}
		& u=0 \quad \text { at } \quad y=\Lambda \\
		& \left\{\begin{array}{ll}
			u=0, & \phi/2<|z| \leqslant 1 / 2 \\
			\partial_y u-\epsilon \left ( \eta \partial_{y y} u+ \eta^{\prime} \partial_z u\right )=0, & |z| \leqslant \phi/2.
		\end{array} \text { at } y=0\right. .
	\end{aligned}
\end{equation}
After substituting the perturbation expansion for the velocity field , $u=u^{(0)}+\epsilon u^{(1)}+O\left(\epsilon^2\right)$, to solve the first-order velocity $\tilde{u}^{(1)}$ (seen Appendix \S \ref{Procedure_lambda_perp_f_c}) as
\begin{equation}
	\tilde{u}^{(1)}=r_0^{(1)} +q_0^{(1)}\frac{y}{\Lambda}+\sum_{n=1}^{\infty}\left[r_n^{(1)} \cosh \left(2\pi n y\right)+q_n^{(1)} \sinh \left(2\pi n y\right)\right] \cos \left(2\pi n z\right),
\end{equation}
we can obtain numerically the coefficients $r_0^{(0)}$, $r_n^{(0)}$, $r_0^{(1)}$ and $r_n^{(1)}$. Consequently, the total flow rate $Q$ can be calculated by
\begin{equation}
	\begin{aligned}
		Q & =\int_{-1/2}^{ 1/2} \int_{0}^{\Lambda} u^{(0)} \mathrm{d} y \mathrm{~d} z+\epsilon\left(\int_{-1/2}^{ 1/2} \int_{0}^{\Lambda}  u^{(1)} \mathrm{d} y \mathrm{~d} z+\int_{-\phi/2}^{\phi/2}  u^{(0)}(0, z) \eta \mathrm{~d} z\right)+O\left(\epsilon^2\right)
		\\ &=Q^{(0)}+\epsilon Q^{(1)}+O\left(\epsilon^2\right),
	\end{aligned}
\end{equation}
where 
\begin{subequations}
    \begin{equation}
	Q^{(0)} = -\frac{dP}{dx}\left[\frac{{\Lambda}^3}{12}+\frac{{\Lambda}^2r_0^{(0)}}{4}\right],
\end{equation}
\begin{equation}
	Q^{(1)} = -\frac{dP}{dx}\left[\frac{{\Lambda}^2r_0^{(1)}}{4}+
	\frac{\phi^3\Lambda}{3}r_0^{(0)}+\Lambda\sum_{n=1}^{\infty} r_n^{(0)} \frac{\sin(n\pi\phi)-n\pi\phi\cos(n\pi\phi)}{n^3\pi^3}  \right].
 \label{Q_first_order}
\end{equation}
\end{subequations}
Based on $Q=-\partial_xP\Lambda^3/12+Q_d  = -\partial_xP\Lambda^3/12 -\partial_xP\Lambda^3\lambda/(4\Lambda+4\lambda)$, the slip length $\lambda_{\parallel,c} $ along the curved interface is 
\begin{equation}
\begin{aligned}
    \lambda_{\parallel,c}
    &= \lambda_{\parallel}^{(0)} + \epsilon \lambda_{\parallel}^{(1)} 
    = \frac{r_0^{(0)}\Lambda}{\Lambda - r_0^{(0)}}
    + \epsilon \frac{4\left(\Lambda + \lambda_{\parallel}^{(0)}\right)^2}{{\Lambda}^4}\frac{Q^{(1)}}{-\partial_X P} = \frac{r_0^{(0)}\Lambda}{\Lambda - r_0^{(0)}} \\
    & 
    + \epsilon \frac{4\left(\Lambda + \lambda_{\parallel}^{(0)}\right)^2}{{\Lambda}^4}\left[\frac{{\Lambda}^2 r_0^{(1)}}{4} +
    \frac{\phi^3 \Lambda}{3} r_0^{(0)} + \Lambda\sum_{n=1}^{\infty} r_n^{(0)} \frac{\sin(n\pi\phi)-n\pi\phi\cos(n\pi\phi)}{n^3\pi^3} \right].
\end{aligned}
\label{total_lambda}
\end{equation}
\subsection{A one-dimensional description  of CCM with slip}\label{average_formulation_ccm}
  The velocity slip length  $\lambda \equiv \lambda^*/l^*$ and temperature slip length $\lambda_{t}\equiv \lambda_t^*/l^*$, describing average effects from the mixed boundary conditions of the two-dimensional problem, enable us to simplify the CCM problem into an effective problem describing flow and heat exchange along the direction of liquid flow, shown in the schematic diagram in Figure \ref{CCM_schematic}f, where 
the notation $ \bar{()} $ implies that the variables are averaged in the $z$-direction.
The corresponding governing equations (\ref{continuity}), (\ref{momentum}), (\ref{temperature}) and (\ref{interface}) can be  replaced, respectively, by 
\begin{subequations}
\begin{equation}
    \partial_X \bar u +\partial_Y \bar v = 0,
    \label{continuity_average}
\end{equation}
    \begin{equation}
0=-\partial_X \bar P+\partial_Y^2 \bar u,
\label{u_bartialP}
\end{equation}
\begin{equation}
0=\partial_Y^2 \bar T,
\label{T_y}
\end{equation}
\begin{equation}
	\partial_Y \bar T(X,h)= \frac{d\Delta}{d \tau}
	\label{interface_average}.
\end{equation}
\end{subequations}
Hence, the corresponding boundary conditions in Figure \ref{CCM_schematic}f are
\begin{subequations}
    \begin{equation}
    \bar u(X,0) = l\lambda \partial_Y \bar u(X,0), \quad \bar u(X,h)=0
    \label{u_bc}
\end{equation}
\begin{equation}
    1-\bar T(X,0)=-l\lambda_{t}\partial_Y \bar T(X,0), \quad  \bar T(X,h)=0
    \label{T_bc}
\end{equation}
\begin{equation}
    \bar v(X,0)=0,\quad \bar v(X,h)=\left[ \left(A-{\mathscr{h}}\right)\frac{\rho}{\mathscr{h}}\frac{d \Delta}{d \tau}+\left(\rho-1\right)\frac{d h}{d \tau}\right]\frac{Ste}{\rho Pe\mathscr{h}\left(A-\mathscr{h}\right)}
    \label{v_bc}
\end{equation}
\begin{equation}
    \partial_X \bar P(0,Y)=0, \quad \bar P(1/2,Y)=0.
    \label{P_bc}
\end{equation}
\end{subequations}
It is worth noting that the upper velocity boundary $\bar v(X,h)$ is derived from $\bar v^*(x^*,h^*) = \rho\partial_{t^*} H^* - \partial_{ t^*} h^*$ based on mass conservation at the melting front. 

The velocity profile $\bar u(X,Y)$ can be obtained by integrating Equation (\ref{u_bartialP}) twice with respect to $Y$ and applying boundary conditions (\ref{u_bc}), yielding 
\begin{equation}
	\bar u = \frac{h^2}{2}\frac{\partial \bar P}{\partial X}\left(\frac{Y^2}{h^2}-\frac{Y/l+\lambda}{\Lambda+\lambda}\right),
	\label{u}
\end{equation}
where $ h/l = \Lambda$.
Then, substituting (\ref{u}) into the continuity equation (\ref{continuity_average}) and integrating along $y$ from 0 to $h$ leads to 
\begin{equation}
	\frac{h^2}{2}\frac{\partial^2 \bar P}{\partial X^2}\int_{0}^{h}\left(\frac{Y^2}{h^2}-\frac{Y/l+\lambda}{\Lambda+\lambda}\right) dY+ \underbrace{\int_{0}^{h}\frac{\partial v}{\partial Y} dY}_{\bar v|_{Y=h}-\bar v|_{Y=0}}=0.
 \label{integral}
\end{equation}
Noting that $\bar v(X,h) \approx Ste \partial_\tau\Delta /(Pe\mathscr{h}^2) $ due to the typical property $\rho-1 < \mathcal{O}(10^{-1})$, then substituting (\ref{v_bc}) into (\ref{integral}) leads to 
\begin{equation}
	\frac{\partial^2 \bar P}{\partial X^2} = \frac{d \Delta}{d\tau} \frac{Ste}{Pe\mathscr{h}^2}\frac{12}{h^3}\frac{\Lambda+\lambda}{\Lambda+4\lambda} .
	\label{pressure second derevative}
\end{equation}
The pressure distribution can be obtained by integrating (\ref{pressure second derevative}) twice with respect to $x$, and using conditions (\ref{P_bc}), to obtain  
\begin{equation}
	\bar P = \frac{d \Delta}{d\tau} \frac{Ste}{Pe\mathscr{h}^2}\frac{12}{h^3}\frac{\Lambda+\lambda}{\Lambda+4\lambda} \left(X^2-\frac{1}{4}\right).
	\label{P}
\end{equation}
Then, we can substitute (\ref{P}) into the force balance (\ref{pressure}) to obtain a relationship between \(\Delta\) and \(h\):
\begin{equation}
	 \frac{d \Delta}{d\tau} = -\frac{Pe\mathscr{h}^2}{Ste}h^3\frac{\Lambda+4\lambda}{\Lambda+\lambda} \Delta^\mathscr{c}.
	\label{Delta-h}
\end{equation}
The temperature profile can be obtained by substituting (\ref{T_bc}) into (\ref{T_y}), yielding
\begin{equation}
	\bar T = \frac{\Lambda-Y/l}{\Lambda+\lambda_{t}}.
	\label{temperature profile}
\end{equation}
 Substitution of (\ref{temperature profile}) into (\ref{interface_average}) gives
\begin{equation}
	\frac{d\Delta}{d\tau}=-\frac{1}{l(\Lambda+\lambda_{t})}.
	\label{Delta-tau}
\end{equation}
The combination of (\ref{Delta-h}) and (\ref{Delta-tau}) leads to
\begin{equation}
	h^4\frac{\Lambda+4\lambda}{\Lambda+\lambda}\left(1+\frac{\lambda_{t}}{\Lambda}\right)=\frac{Ste}{Pe \mathscr{h}^2}\frac{1}{\Delta^\mathscr{c}}.
	\label{Delta}
\end{equation}
Equations (\ref{Delta-tau}) and (\ref{Delta}) are coupled expressions between the solid remaining height $\Delta$ and film thickness $h$. Note that because of the definition $\Lambda=h/l$ where $l$ is fixed, $h$ varies with $\tau$.

Recalling $\mathscr{h}=h_0^*/L^*$ and $h=h^*/h_0^*$ in (\ref{nondimensional}) indicates that $h_0^*$ has not been well determined.
Usually, the initial height satisfies \( H_0^* \gg h^* \) (i.e., $\Delta \approx H$) and we can further find a characteristic initial film thickness $h_0^*$ satisfies the $Ste/Pe\mathscr{h}^2=1$ corresponding to initial solid height $H=1$ and no-slip surface $\lambda=\lambda_t=0$, enabling us to define $h_0^*$ as
\begin{equation}
    h_0^* = 
      \left[\frac{(T_w^*-T_m^*)k_l^*{L^*}^2\mu^*}{\mathcal{H}^*\rho_l^*p_c^*}\right]^{\frac{1}{4}}, 
      \label{h_0_star}
\end{equation}
which is able to transform (\ref{Delta}) and (\ref{Delta-tau}) into the following equations ($\Lambda=h/l$)
\begin{subequations}
\begin{equation}
    h^4\frac{\Lambda+4\lambda}{\Lambda+\lambda}\left(1+\frac{\lambda_{t}}{\Lambda}\right) = \frac{1}{H^\mathscr{c}} \label{H-h},
\end{equation}
\begin{equation}
    \frac{dH}{d\tau} = -\frac{1}{l(\Lambda+\lambda_{t})}.  \label{H-tau}
\end{equation}
\end{subequations}
Based on the equation (\ref{H-h}), we can conclude that a constant film thickness \(h\) is maintained for the constant-pressure mode ($\mathscr{c}=0$) though $\lambda$ and $\lambda_t$ are dependent on $h$. On the other hand, for the gravity-driven mode ($\mathscr{c}=1$)  there is a coupling  between \(h\) and \(H\), making the solutions of $H(\tau)$ and  $h(\tau)$ complicated. 

Furthermore, we introduce the Nusselt number, based on the instantaneous height of the liquid film $h^*(t^*)$, to demonstrate the heat transfer capability under different conditions, which is defined as
\begin{equation}
	Nu \equiv -\frac{k_l^{*}\partial_{y^{*}} \bar T^{*}(0,h^{*})}{\left(T_w^{*}-T_m^{*}\right)}\frac{h^{*}}{k_{l}^{*}} = \frac{1}{l(\Lambda+ \lambda_{t})}.
	\label{Nu}
\end{equation}

At this stage, we have completed all the analysis of the thin film flow and heat transfer necessary for modeling CCM processes, enhanced using a corrugated substrate. The main variables that dictate performance are the slip lengths $\lambda$ and $\lambda_t$, geometric ratios $l$ and $\Lambda$, the latter which changes in time, and the resulting $Nu$, which determines the melting rate, i.e., the power density of the CCM.

\section{Results and discussions}
\label{sec_3}
The goal of this section is to calculate the melting rate (i.e., $Nu$) or the instantaneous solid height (i.e., $H(\tau)$), which is a complicated function of geometric parameters ($l$) and slip lengths ($\lambda$, $\lambda_t$). It also depends on the operation controlled by hydrostatic pressure or a constant applied pressure. The different subsections summarize the main results for different cases.
\subsection{Confinement and meniscus effects on the slip length} \label{sec_slip}
\subsubsection{Velocity slip length $\lambda$ and temperature slip length $\lambda_t$ at a flat interface}
\label{asym_flat}
\begin{figure}
\captionsetup{justification=justified}
	\centerline{\includegraphics[width=0.9\textwidth]{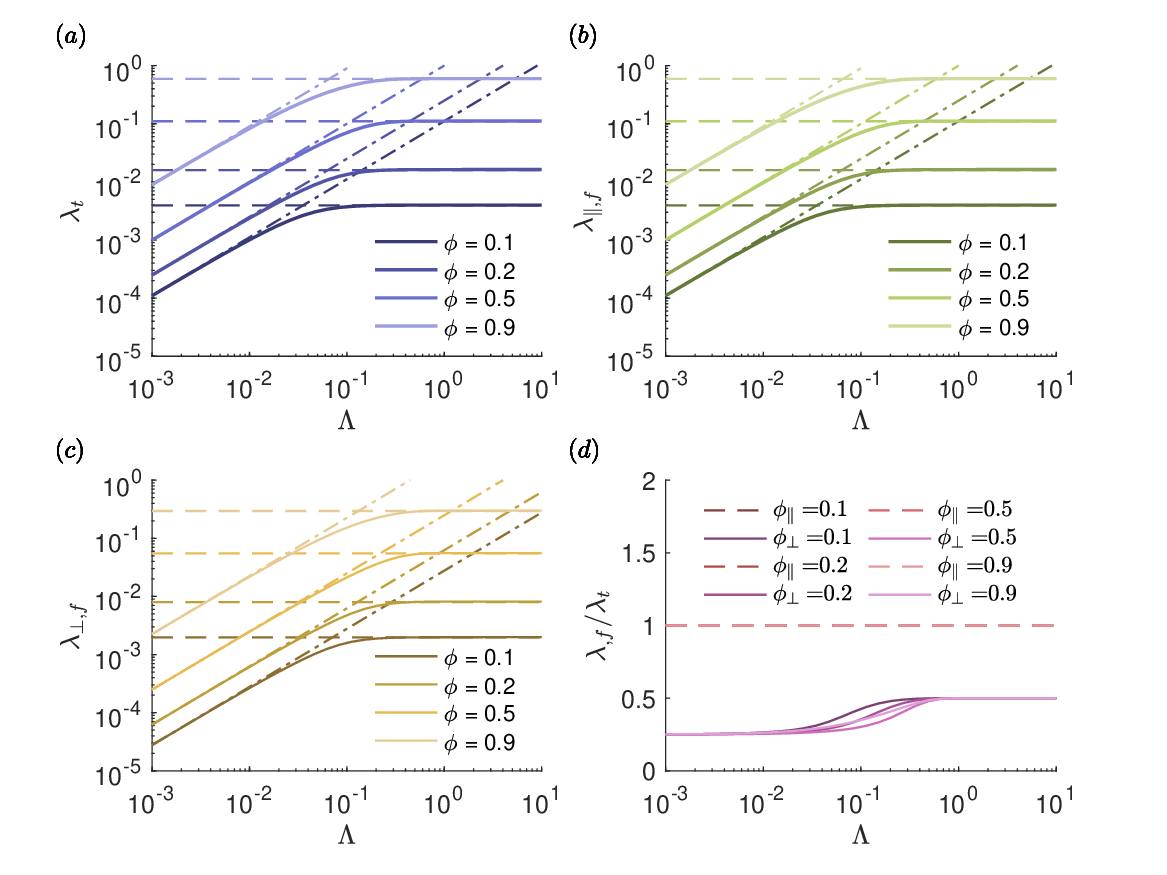}} 
	\caption{Slip lengths as a function of $\Lambda$. (a) Temperature slip length $\lambda_t$, velocity slip length (b) $\lambda_{\parallel,f}$ on longitudinal grooves and (c) $\lambda_{\perp,f}$ on transverse grooves over aspect ratio $\Lambda=h^*/l^*$ when liquid-gas interface is flat. (d) Comparison of velocity-temperature slip length $\lambda_{(),f}/\lambda_t$ between longitudinal and transverse grooves. In figures (a-c) solid lines are numerical results. Dashed lines and dot-dash lines are asymptotic solutions
	}
	\label{flat_slip_length}
\end{figure}
The temperature $\lambda_t$ (\S \ref{sec_lambda_t_f}) and velocity $\lambda_{\parallel,f}$ slip lengths (\S \ref{sec_slip_length_f}) as a function of the aspect ratio $\Lambda=h^*(t^*)/l^*$ (see Figure \ref{CCM_schematic}a), for a given gas-liquid fraction $\phi$, are presented as solid lines in Figures \ref{flat_slip_length}a and \ref{flat_slip_length}b, respectively. For comparison, we also calculate the slip length $\lambda_{\perp,f}$ variation on a surface of transverse grooves in Appendix \ref{lambda_perp}, and present it in Figure \ref{flat_slip_length}c. It can be observed that across a wide range of gas-liquid fractions \(\phi\), each slip length of $\lambda_t$, $\lambda_{\parallel,f}$, and $\lambda_{\perp,f}$  maintains a specific fixed value for \(\Lambda  \gg 1 \), and exhibits a scaling law of \(\sim \Lambda\) when \(\Lambda \ll 1\). 

Asymptotic solutions for these slip lengths are readily available, which in Figure \ref{flat_slip_length} are drawn as dashed lines for $\Lambda \gg 1$ and dot-dash lines for $\Lambda \ll 1$. In the limit of $\Lambda \gg 1$, the asymptotic solutions of $\lambda_{\parallel,f}$ and $\lambda_{\perp,f}$ are well known \citep{RN1091,RN1046} as 
\begin{subequations}
    \begin{equation}
\lambda_{\parallel,f}= \frac{1}{\pi} \ln \left(\sec \left(\frac{\phi \pi}{2}\right)\right), \quad \Lambda \gg 1,
\label{slip_para_inf}
\end{equation}
\begin{equation}
\lambda_{\perp,f}= \frac{1}{2\pi} \ln \left(\sec \left(\frac{\phi \pi}{2}\right)\right), \quad \Lambda \gg 1.
\end{equation}
\end{subequations}
In the limit of $\Lambda \ll 1$, we found the asymptotic solutions proposed by \citet{RN1046} are incorrect due to mistakes in calculating $Q_d$. The velocity slip length $\lambda_f$ should be 
\begin{subequations}
    \begin{equation}
	\lambda_{\parallel,f}= \frac{\phi}{1-\phi}\Lambda, \quad \Lambda \ll 1,
\label{lambda_para_0}
\end{equation}
\begin{equation}
	\lambda_{\perp,f}= \frac{\phi}{4\left(1-\phi\right)}\Lambda, \quad \Lambda \ll 1,
\end{equation}
\end{subequations}
which are consistent with the other related results \citep{RN1091,RN1048}. Furthermore,  asymptotic solutions of the temperature slip length are obtained as  
\begin{subequations}
    \begin{equation}
	\lambda_{t} = \frac{1}{\pi} \ln \left(\sec \left(\frac{\phi \pi}{2}\right)\right), \quad \Lambda \gg 1,
 \label{slip_temp_inf}
\end{equation}
\begin{equation}
	\lambda_{t} = \frac{\phi}{1-\phi}\Lambda , \quad \Lambda \ll 1.
\label{lambda_t_0}
\end{equation}
\end{subequations}
As already highlighted at the end of \S \ref{sec_slip_length_f}, based on the results reported in Figures \ref{flat_slip_length}a and b, it is observed that \(\lambda_{\parallel, f}=\lambda_t\), while \(\lambda_{\perp, f}\) remains a constant value proportional to them in both asymptotic ranges. As shown in Figure \ref{flat_slip_length}d,  the ratio \(\lambda_{\perp,f}/\lambda_t\) varies between 0.25 and 0.5 depending on  \(\phi\) and \(\Lambda\), consistent with \(\lambda_{\parallel,f} = \lambda_t = 2\lambda_{\perp,f} = \ln\left(\sec(\phi\pi/2)\right)/\pi\) for $\Lambda \gg 1$, and \(\lambda_{\parallel,f} = \lambda_t = 4\lambda_{\perp,f} = \phi \Lambda/(1-\phi)\) for $\Lambda\ll 1$.

\subsubsection{Velocity slip length $\lambda_{\parallel,c}$ at a curved gas-liquid interface}

\begin{figure}
\captionsetup{justification=justified}
\centerline{\includegraphics[width=0.9\textwidth]{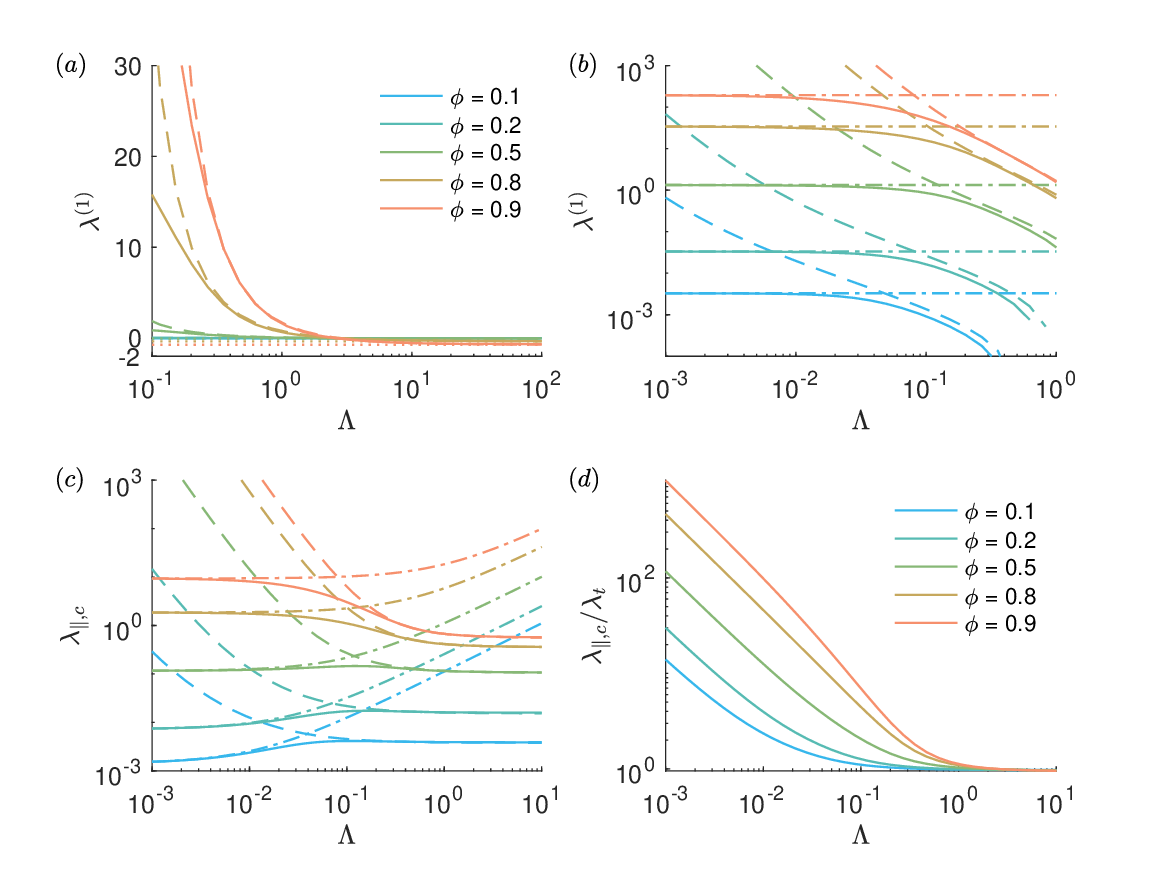}} 
	\caption{Slip lengths at a curved interface. (a-b) Variation of the first-order velocity slip length $\lambda^{(1)}$ for different $\Lambda$, where a dotted line denotes (\ref{old-lambda_infty}), a dashed line (\ref{new lambda_infty}), and a dot-dash line (\ref{lambda_0}). (c) Total velocity slip length $\lambda_{\parallel,c}=\lambda^{(0)}+\epsilon\lambda^{(1)}$ versus $\Lambda$, with a dashed line denoting (\ref{total_lambda_curved}) and a dot-dash line  (\ref{total_lambda_curved_0}). (d) The ratio of the total velocity and temperature slip lengths, $\lambda_{\parallel,c}/\lambda_t$, versus $\Lambda$ .
	}
	\label{first_order_slip}
\end{figure}

We next consider velocity slip length $\lambda_{\parallel,c}$ at a curved gas-liquid interface as a function of $\Lambda$ and $\phi$. 
The semi-log and log-log results of the first-order slip length $\lambda^{(1)}$, computed as described in the \S \ref{sec_slip_length_c}, are shown in Figure \ref{first_order_slip}a and b, respectively. For different \(\phi\), \(\lambda^{(1)}\) exhibits two plateau intervals as \(\Lambda \ll 1\)  and \(\Lambda \gg 1\), with a decreasing trend along with $\Lambda$ in between. Notably, when \(\Lambda \gtrapprox 2\) , \(\lambda^{(1)}\) becomes negative. This occurs because the increased area for flow caused by the meniscus is negligible, while there is a reduced velocity gradient at the boundary $y=0$ due to the meniscus, which results in a negative slip length.

The asymptotic solution of the first-order slip length $\lambda^{(1)}$, with $\mathcal{O}(\Lambda^{-1})$ error,  has already been proposed \citep{RN1096,RN1062},
\begin{equation}
	\lambda_{\|}^{(1)}= -\phi^3\int_{0}^{1} \frac{\left[1-\cos\left(\phi \pi s\right)\right]\left(1-s^2\right)}{\cos\left(\phi\pi s\right)-\cos\left(\phi \pi\right)}\mathrm{d}s + \mathcal{O}(\Lambda^{-1}) , \quad \Lambda \gg 1,
	\label{old-lambda_infty}
\end{equation}
which only accounts for the changed velocity profile and neglects effects of increased flow rate associated with an expanded area, corresponding to the slip length 
\begin{equation}
\lambda_{\|, c}=\ln (\sec (\phi \pi / 2)) / \pi-\epsilon \phi^3 \mathcal{F}(\phi),\quad \Lambda \gg 1;
\label{total_lambda_curved_gg_1}
\end{equation}the function $\mathcal{F}(\phi)$ is defined shortly. 
Based on the (\ref{Q_first_order}) in the limit of $\Lambda \gg 1$, we can give the flow rate $Q^{(1)}$ as
\begin{equation}
	Q^{(1)}= -\frac{\phi^3\Lambda^2l}{4} \mathcal{F}(\phi)
	+\pi\phi^4\Lambda l^2	\mathcal{G}(\phi),\quad \Lambda \gg 1,
\end{equation}
where $\mathcal{F(\phi)}$ and $\mathcal{G(\phi)}$ are
\begin{equation}
	\mathcal{F}(\phi)=\int_{0}^{1} \frac{\left[1-\cos\left(\phi \pi s\right)\right]\left(1-s^2\right)}{\cos\left(\phi\pi s\right)-\cos\left(\phi \pi\right)}\mathrm{d}s
\end{equation}
and
\begin{equation}
	\mathcal{G}(\phi)=\int_{0}^{1} \frac{s\left(1-s^2/3\right)\sin\left(\phi\pi s/2\right)}{\sqrt{\cos\left(\phi\pi s\right)-\cos\left(\phi \pi\right)}}\mathrm{d}s.
\end{equation}

\begin{table}
	\newcommand{\tabincell}[2]{\begin{tabular}{@{}#1@{}}#2\end{tabular}}
	\centering
	\setlength{\tabcolsep}{2.5mm}{
		\begin{tabular}{cccccc}
			
			\text{interface} & slip length          & $h^*/l^* \gg 1$   & $h^*/l^* \ll 1$ \\
			
			\\
			flat ($\theta=0$)      &     $ \lambda_{\parallel}^{*}$           & $\frac{l^{*}}{\pi} \ln \left(\sec \left(\frac{\phi \pi}{2}\right)\right)$                     & $\frac{\phi}{1-\phi}h^{*} $                  \\
			
			flat ($\theta=0$)      & $ \lambda_{\perp}^{*}$                      & $\frac{l^{*}}{2\pi} \ln \left(\sec \left(\frac{\phi \pi}{2}\right)\right)$            &$\frac{\phi}{4(1-\phi)}h^{*}$                           \\
			
			curved or flat    & $ \lambda_{t}^{*}$                     & $\frac{l^{*}}{\pi} \ln \left(\sec \left(\frac{\phi \pi}{2}\right)\right)$                     & $\frac{\phi}{1-\phi}h^{*}$                      \\
			curved ($\theta>0$)      & $ \lambda_{\parallel}^{*}$    &$\frac{l^{*}}{\pi} \ln \left(\sec \left(\frac{\phi \pi}{2}\right)\right)-\epsilon\phi^3l^{*}\mathcal{F}(\phi)$                  &$\frac{\phi}{1-\phi}h^{*} +\epsilon\frac{8\phi^3}{3\left(1-\phi\right)^2}l^{*}$                      \\   

\\
interface & slip length & \multicolumn{2}{c}{$h^*/l^* \gtrapprox 0.2$} \\ 
\\
curved ($\theta>0$) & $\lambda_{\parallel}^{*}$ & \multicolumn{2}{c}{$\frac{l^{*}}{\pi} \ln \left(\sec \left(\frac{\phi \pi}{2}\right)\right) + l^{*}\epsilon\left(-\phi^3\mathcal{F}(\phi) + \frac{4l^{*}}{h^{*}}\phi^4\mathcal{G}(\phi)\right)\left(1 + \frac{l^{*}}{4h^{*}\pi}\ln\left(\sec\left(\frac{\phi\pi}{2}\right)\right)\right)^2$} \\ 
\\

\\
	\end{tabular}}
	\caption{Dimensional asymptotic formula of velocity ($\lambda_{\parallel}^{*}$ or $\lambda_{\perp}^{*}$) and thermal ($\lambda_{t}^{*}$) slip length for  various confinement and meniscus effects.}
	\label{asymptotic}
\end{table}
Hence, a modified asymptotic solution of $\lambda_{\parallel}^{(1)}$ for $\Lambda \gtrapprox 0.2$ is proposed as
\begin{equation}
	\lambda_{\|}^{(1)}=\left(-\phi^3 \mathcal{F}(\phi)
	+\frac{4}{\Lambda}\phi^4	\mathcal{G}(\phi)\right)\left(1+
	\frac{1}{4\Lambda\pi} \ln \left(\sec \left(\frac{\phi \pi}{2}\right)\right)\right)^2,\quad \Lambda \gtrapprox 0.2.
	\label{new lambda_infty}
\end{equation}
In the limit of $\Lambda \ll 1$, the asymptotic solution can be derived by substituting $\lambda_{\parallel}^{(0)}=l\phi \Lambda/(1-\phi)$ and $Q^{(1)} = 2\phi^3\Lambda^2l/3$ into the second term of (\ref{total_lambda}), leading to 
\begin{equation}
	\lambda_{\parallel}^{(1)}=\frac{8\phi^3}{3\left(1-\phi\right)^2},\quad \Lambda \ll 1.
	\label{lambda_0}
\end{equation}
As depicted in Figure \ref{first_order_slip}a and b, the asymptotic solutions (\ref{old-lambda_infty}) and (\ref{lambda_0})  for the plateaus can accurately predict the first-order velocity slip length $\lambda^{(1)}$ for different \(\phi\). However, their applicability is limited ($\Lambda \lessapprox 0.01$ or $\Lambda \gtrapprox 10^2$ ), while the newly proposed modified asymptotic solution (\ref{new lambda_infty}) can accurately predict the first-order velocity slip for \(\Lambda \gtrapprox 0.2\).

Based on the velocity slip length of zero-order $\lambda^{(0)}$ and first-order $\lambda^{(1)}$, the integrated slip length \(\lambda_{\parallel,c}\), considering the curved liquid surface, can be finally obtained according to the equation $\lambda_{\parallel,c}=\lambda^{(0)}+\epsilon\lambda^{(1)}$, which is plotted for different $\phi$ as solid lines in Figure \ref{first_order_slip}c. The corresponding asymptotic solutions of $\lambda_{\parallel,c}$ are also drawn as dot-dash and dashed lines, respectively corresponding to the equations 
\begin{equation}
    \lambda_{\parallel,c}=\frac{\phi}{1-\phi}\Lambda+
    \epsilon\frac{8\phi^3}{3\left(1-\phi\right)^2},\quad \Lambda \lessapprox 0.01
 \label{total_lambda_curved_0}
\end{equation}
\begin{equation}
    \lambda_{\parallel,c}=\frac{1}{\pi} \ln \left(\sec \left(\frac{\phi \pi}{2}\right)\right)+
    \epsilon\left(-\phi^3 \mathcal{F}(\phi)
	+\frac{4}{\Lambda}\phi^4	\mathcal{G}(\phi)\right)\left(1+
	\frac{1}{4\Lambda\pi} \ln \left(\sec \left(\frac{\phi \pi}{2}\right)\right)\right)^2,\quad \Lambda \gtrapprox 0.2.
 \label{total_lambda_curved}
\end{equation}
The presence of meniscus leads to a constant value of $\lambda_{\parallel,c}$ when $\Lambda \ll 1$.  Hence, the results of $\lambda_{\parallel,c}/\lambda_t$ indicate a considerable value when $\Lambda \ll 1$ for arbitrary $\phi$ as plotted in Figure \ref{first_order_slip}d.
The asymptotic solutions obtained in this section \S \ref{sec_slip} are summarized in Table \ref{asymptotic} for convenient comparison.

\subsection{Characterizing constant-pressure close-contact melting} \label{constant_pressure_ccm}
We now use the main results of \S \ref{sec_slip} to calculate the melting rate.
Recalling (\ref{H-h}) with $\mathscr{c}=0$, we observe that a constant dimensionless liquid film thickness \(h=h^*/h^*_0\) can be determined, yielding 
\begin{equation}
	h = \left[\frac{1+ \lambda/\Lambda}{\left(1+4 \lambda/\Lambda\right)\left(1+ \lambda_{t}/\Lambda\right)}\right]^{\frac{1}{4}}.
	\label{h_constant_P}
\end{equation}
Substitution of (\ref{h_constant_P}) into the definition of $Nu$ (\ref{Nu}) leads to a prediction for the melting rate
\begin{equation}
	Nu = \left[\frac{1+4\lambda/\Lambda}{\left(1+ \lambda/\Lambda\right)\left(1+ \lambda_{t}/\Lambda\right)^3}\right]^{\frac{1}{4}}.
 \label{mathcal_K_eq}
\end{equation}
Note that \(Nu = 1\) represents the heat transfer performance of a smooth, unstructured surface ($\lambda=\lambda_t=0$). Therefore, \(Nu > 1\) indicates that the effects of velocity slip and temperature slip enhance heat transfer. Conversely, \(Nu < 1\) implies a reduction in heat transfer. 

\subsubsection{Flat liquid-gas interface}
\begin{figure}\captionsetup{justification=justified}\centerline{\includegraphics[width=\textwidth]{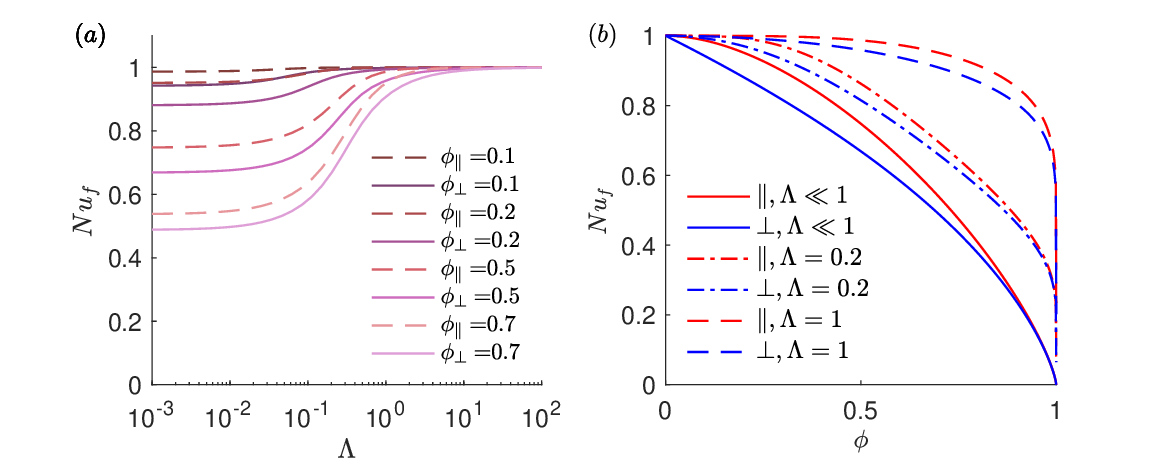}} 
	\caption{  Variation of $Nu_f$ along with (a) aspect ratio $\Lambda$ by numerical results and (b) liquid-gas fraction $\phi$ by asymptotic formula for flat gas-liquid interface. $\parallel$ and $\perp$ represent longitudinal and transverse grooves respectively.
	}
	\label{mathcal_K}
\end{figure}
Substituting the numerical results of \(\lambda_{\parallel,f}\) and \(\lambda_{\perp,f}\) respectively obtained from \S \ref{sec_lambda_t_f} and Appendix \ref{lambda_perp} into Equation (\ref{mathcal_K_eq}), yields the results of $Nu_f$ versus $\Lambda$ for a flat interface shown in Figure \ref{mathcal_K}a. For both longitudinal and transverse groove structures, the existence of slip always reduces heat transfer. When \(\Lambda \gg 1\) , \(Nu_f\) asymptotically equals 1 due to negligible slip effect. For the case of \(\Lambda \ll 1\), \(Nu_f\) decreases and then asymptotically equals a constant value for either the longitudinal or transvers grooves. It is obvious that \(Nu_{\parallel,f} > Nu_{\perp,f}\) is valid for any \(\phi\). Therefore, we can determine that \(Nu_f\) varies only as a function of \(\phi\) within the the asymptotic region (\(\Lambda \ll 1\) as indicated in Figure \ref{flat_slip_length}). By substituting the asymptotic solutions in Section \ref{asym_flat}, we find 

\begin{equation}
    Nu_{\parallel,f}=\left[\left(1+3\phi\right)\left(1-\phi\right)^3\right]^{\frac{1}{4}}, \quad \Lambda \ll 1;
\end{equation}
\begin{equation}
    Nu_{\perp,f}=\left[\frac{4\left(1-\phi\right)^3}{4-3\phi}\right]^{\frac{1}{4}},\quad \Lambda \ll 1,
\end{equation}
which are plotted in Figure \ref{mathcal_K}b. All curves generally show a decreasing trend of \(Nu\) as \(\phi\) increases. For \(\Lambda \ll 1\), the parallel and perpendicular values of \(Nu\) start at 1 and decrease smoothly. For larger values of \(\Lambda = \) 0.2 and 1, the curves for both \(Nu_{\parallel,f}\) and \(Nu_{\perp,f}\) exhibit more pronounced decreases and diverge more significantly as \(\phi\) approaches 1. The curves suggest that the parameter \(\Lambda\) significantly impacts \(Nu_f\), with higher values of \(\Lambda\) leading to more rapid decreases in \(Nu_f\).

\subsubsection{A curved liquid-gas interface on longitudinal grooves}
\label{cureved_const_pressure}
\begin{figure}\captionsetup{justification=justified}\centerline{\includegraphics[width=\textwidth]{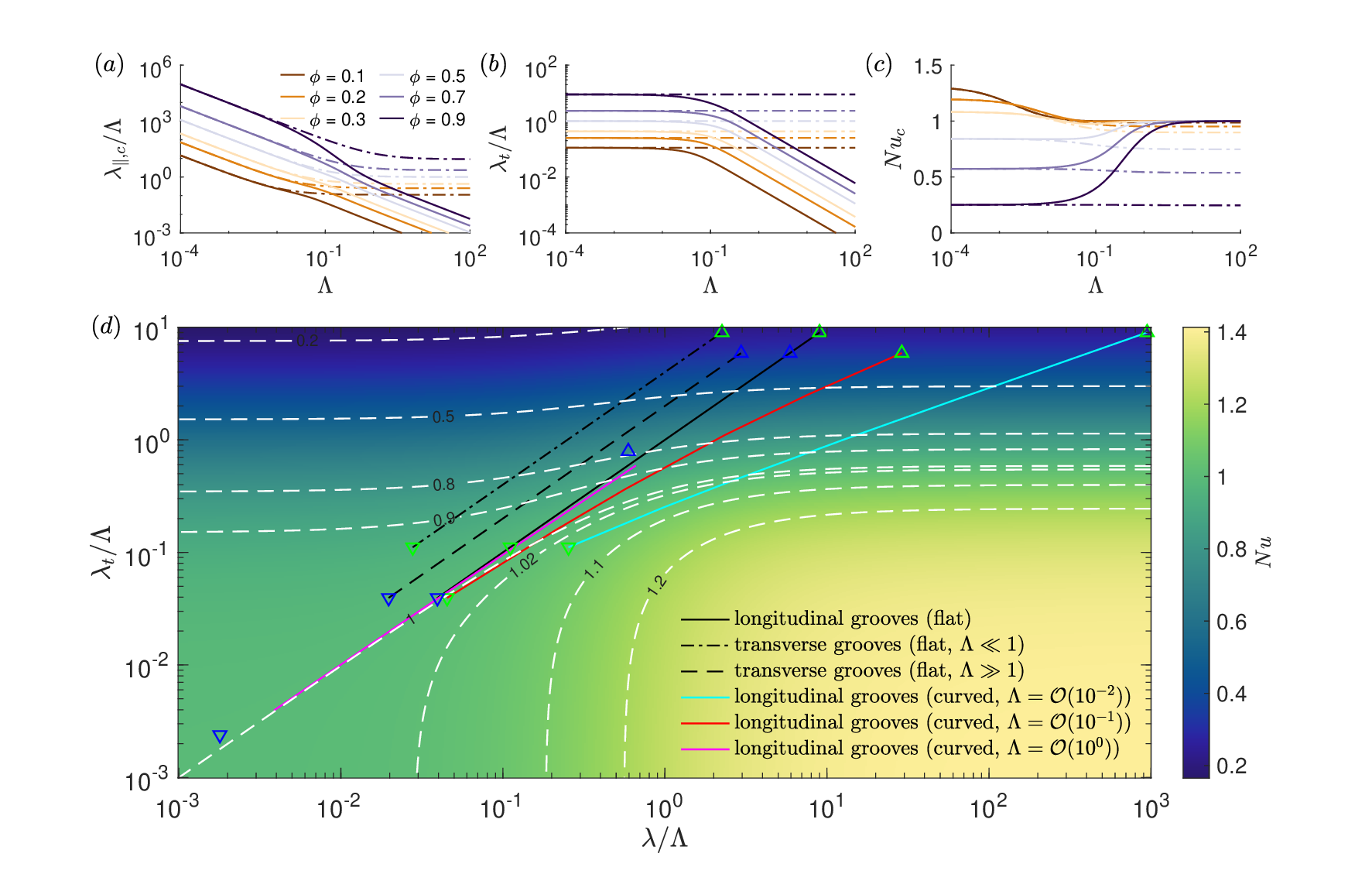}} 
	\caption{ Variation of (a) \(\lambda/\Lambda\) and (b) \(\lambda_t/\Lambda\) versus \(\Lambda\), where solid lines represent numerical results and dot-dash lines represent asymptotic solutions listed in Table \ref{asymptotic}. (c) Variation of $Nu$ versus $\Lambda$ for curved gas-liquid interface, dot-dash lines are asymptotic solutions (\ref{Nu_c_0}). (d) Map of $Nu$ at the various combinations of $\Lambda$, $\phi$ and transverse/longitudinal surface structures for constant-pressure mode. Symbols denote {\textcolor{blue}{$\bigtriangledown$}}: $\phi=0.1$ and $\Lambda \gg 1$, {\textcolor{blue}{$\triangle$}}: $\phi=0.9$ and $\Lambda \gg 1$, 
 {\textcolor{green}{$\bigtriangledown$}}: $\phi=0.1$ and $\Lambda \ll 1$,
  {\textcolor{green}{$\triangle$}}: $\phi=0.9$ and $\Lambda \ll 1$.
	}
	\label{K-map}
\end{figure}

For a curved liquid-gas interface on longitudinal grooves, we firstly observe that \(\lambda_{\parallel,c}/\Lambda\) and \(\lambda_t/\Lambda\) can be considered as the functions of \(\Lambda\) as shown in Figures \ref{K-map}a and b, respectively. 
Similarly, by substituting the numerical solutions for $\lambda_{\parallel,c}$ and $\lambda_{t}$ into the equation (\ref{mathcal_K_eq}), we can obtain the numerical results of $Nu_c$ for considering meniscus on longitudinal grooves, which is plotted in the Figure \ref{K-map}c. Compared to the flat interface, \(Nu_c\) is also close to 1 for $\Lambda \gg 1$. However, as \(\Lambda \to 0\), there exists a critical \(\phi\) at which \(Nu_c\) can  be larger than 1 and heat transfer is enhanced. By substituting the asymptotic slip length solution (\ref{total_lambda_curved_0}) at \(\Lambda \ll 1\) into (\ref{mathcal_K_eq}), we find
\begin{equation}
Nu_{c}=\left(\frac{(1-\phi)^3\left(3 \Lambda\left(1+2 \phi-3 \phi^2\right)+8 \phi^2 \sin{\theta}\right)}{3 \Lambda(1-\phi)+2 \phi^2 \sin{\theta}}\right)^{\frac{1}{4}} \overset{\Lambda \rightarrow 0}{=}\left[4(1-\phi)^{3}\right]^{\frac{1}{4}},
\label{Nu_c_0}
\end{equation}
which is plotted as  dot-dash lines for different $\phi$ in Figure \ref{K-map}c. Then, by letting (\ref{Nu_c_0}) =1, we can obtain the critical fraction \(\phi_{cr} = 1-2^{-2/3} \approx 0.37\). Therefore, we can infer that when \(\phi > 0.37\), heat transfer will be reduced regardless of the value of \(\Lambda\). However, when \(\phi < 0.37\), heat transfer can be enhanced to accelerate melting, with \(Nu_c\) approaching a constant value of \(4^{1/4}(1-\phi)^{3/4}\) as \(\Lambda\) decreases.

In order to visually compare the  effects of surface structure, a meniscus, aspect ratio (\(\Lambda\)), and gas-liquid ratio (\(\phi\)) on the \(Nu\), a two-dimensional phase diagram is presented in Figure \ref{K-map}d. In this diagram, the contours correspond to varying \(Nu\), with white dashed lines representing \(Nu\) isocline and other lines indicating results derived from asymptotic solutions. It can be observed that for a flat interface, the \(Nu<1\) always remains for surfaces with transverse grooves structures, indicating a reduction in heat transfer effectiveness. In contrast, surfaces with longitudinal grooves maintain \(Nu \rightarrow 1\) for small \(\phi\) values, but \(Nu\) decreases monotonically as \(\phi\) increases. 
When a meniscus (curved interface) is introduced, the \(Nu>1\) can be achieved for longitudinal grooves initially at small $\phi$ and $\Lambda<1$  and $Nu$ shows a non-monotonous trend versus \(\phi\). The enhancement of $Nu$ at $\phi<0.37$ is more pronounced as \(\Lambda\) decreases.

Overall, under the condition of constant pressure, longitudinal grooves ($\parallel$) exhibit the best heat transfer performance among all surfaces. Furthermore, only with the presence of  a meniscus  and significant confinement effects ($\Lambda \ll 1$), can one achieve an enhancement in heat transfer at an appropriate gas-liquid fraction ($\phi<0.37$).

\subsection{Gravity-driven close-contact melting }
 In contrast with the constant pressure mode just discussed, when the phase change material is supported only by hydrostatic pressure, i.e., the gravity-driven mode (i.e., $\mathscr{c}=1$), the film thickness $h^*(t^*)$ changes in time.
\subsubsection{Numerical results of solid height $H(\tau)$ and film thickness $h(\tau)$}

\begin{figure}
\captionsetup{justification=justified}
\centerline{\includegraphics[width=\textwidth]{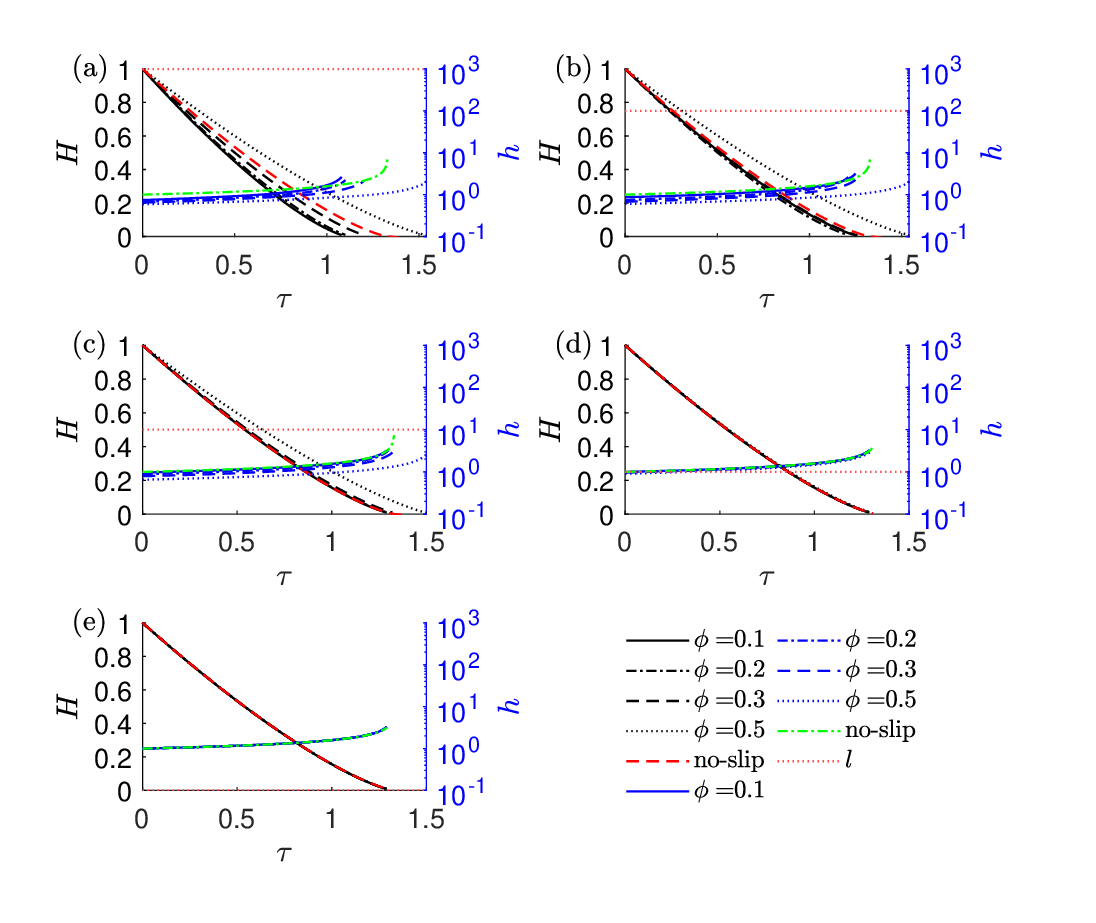}} 
	\caption{ Variation of \(H\) and \(h\) versus $\tau$ for different conditions of (a) $l=10^3$, (b) $l=10^2$, (c) $l=10$ (d) $l=1$ and (e) $l=10^{-1
}$. In all figures, the black line represents the remaining height \(H\), the blue line represents the film thickness \(h\), the red dotted line (\reddotline) represents the magnitude of \(l\), red dashed line (\reddashline) and green dot-dash line(\greendashdotline)  represent  the no-slip solution of $H(\tau)$ (\ref{H_tau_no_slip}) and $h(\tau)$ (\ref{h_tau_no_slip}), respectively. Four values of $\phi= \{0.1, 0.2, 0.3, 0.5\}$ are chosen for calculating each case.
 }
	\label{H_h_t}
\end{figure}

Recalling the equations (\ref{H-h}) and (\ref{H-tau}) with $\mathscr{c}=1$,
we can rewrite them by using $h(\tau)=l\Lambda(\tau)$ from (\ref{nondimensional}), yielding
\begin{subequations}
    \begin{equation}
    \Lambda^4\frac{1+4\lambda/\Lambda}{1+\lambda/\Lambda}\left(1+\frac{\lambda_{t}}{\Lambda}\right) =\frac{1}{Hl^4},
    \label{H_Lambda}
\end{equation}
\begin{equation}
	\frac{dH}{d\tau}=-\frac{1}{l(\Lambda+\lambda_{t}).} 
\end{equation}
\end{subequations}
This set of equations consists of a first-order ODE and an algebraic equation, which can be numerically solved as detailed in Appendix \S \ref{solving_H_tau}.
For comparison, the analytical solution for a no-slip condition can be obtained by substituting $\lambda=\lambda_t=0$ into the above equations, yielding
\begin{subequations}
    \begin{equation}
    H_{no-slip} = \left(1-\frac{3}{4}\tau \right)^{\frac{4}{3}},
\label{H_tau_no_slip}
\end{equation}
\begin{equation}
    h_{no-slip} = \left(1-\frac{3}{4}\tau \right)^{-\frac{1}{3}}.
    \label{h_tau_no_slip}
\end{equation}
\end{subequations}
\\
We estimated that the magnitude of \(l \)  ranges from \(10^{-1}\) to \(10^{3}\), based on the representative values of the following dimensionless parameters: \(Ste = 0.01\)–\(0.1\), \({A}^2 = 0.1\)–\(10\),  and \(H_0^{*}\) varies from 0.01 m to 1 m and \(l^{*}\) varies from 10 $\mu$m to 1 mm, with representative PCM chosen as water and tetradecanol \citep{RN1047,RN386}. It is worth noting that liquid metals are not considered for the PCM due to the inability to meet the lubrication assumptions. The results for $l \in \{10^{3}, 10^{2},10^{1},10^{0}, 10^{-1}\}$ are drawn in Figure \ref{H_h_t} from (a) to (e), respectively. 
Since we have proved that a flat interface ($\parallel,f$) always reduces the melting rate in the \S \ref{constant_pressure_ccm} and it is similar for the gravity-driven CCM here, only results considering a meniscus ($\parallel,c$) are discussed in the following.

For large \(l=10^3\) and \(10^2\), the results show that the melting rate $dH/d\tau$ decreases with increased \(\phi\) (black line) as illustrated  in Figure \ref{H_h_t}a and \ref{H_h_t}b, though the film thickness \(h\) also decreases along with larger \(\phi\) (blue line). This indicates that the temperature slip is more pronounced than the velocity slip. Compared to the case of no-slip (red dashed line),  \(\phi \le 0.3\) exhibits an accelerated melting rate due to slip, similar to the pressure-driven CCM in Figure \ref{K-map}c. 
When $l=10$ in  Figure \ref{H_h_t}c , it can be demonstrated that the slip effects will result in a reduction in the melting rate. Furthermore, the reduction is more pronounced when the value of \(\phi\) is larger, whereas a smaller value of \(\phi\) is close  to the results of the no-slip case.
This trend is also observed when $l=1$ and 0.1, and it can be shown that as $l$ decreases, all slip cases gradually approach those of no-slip (Figure \ref{H_h_t}d and e).

In general, an increase in the value of \(\phi\) will invariably result in a reduction in the thickness of the liquid film \(h\). However, the melting rate is decreased due to the more pronounced effect of temperature slip. The potential for slip to accelerate the melting process depends on the specific combination of \(l\) and \(\phi\). When \(l\) is sufficiently large (e.g., \(10^2\) or \(10^3\)), even a modest increase in \(\phi\) can enhance heat transfer, thereby expediting the melting process. Conversely, as \(l\) diminishes, the influence of slip decreases correspondingly, ultimately becoming negligible.

\subsubsection{Phase diagram and asymptotic solution}

To comprehensively illustrate the effects of \(\phi\) and \(l\), we selected 288 and 110 values from the ranges \(\phi\) from 0.1 to 0.9 and \(l\) from $10^{-2}$ to $10^{3}$, respectively. These values were used to construct the parameter matrix (log$_{10}l$, $\phi$), which were then substituted into the numerical solution calculations to obtain the total melting time $\tau_{end}$ defined as $H(\tau_{end})=0$ for each case of (log$_{10}l$, $\phi$). The resulting phase diagram is shown in Figure \ref{diagram}, where the contour represents \(\tau_r = \tau_{end}/\tau_{no-slip}=3\tau_{end} /4 \), meaning the melting time ratio of slip surface to no-slip surface, where $\tau_{no-slip}=4/3$ from (\ref{H_tau_no_slip}). Therefore, $\tau_r <1$ means heat transfer enhancement, while $\tau_r >1$ means heat transfer reduction. 

It can be observed that the region where enhancement effects can be achieved exists only in the lower right corner of the phase diagram. The upper limit boundary is at \(\phi = 1 - 2^{-2/3}\), which is consistent with the conclusions drawn from the constant pressure CCM in \S \ref{cureved_const_pressure}. The left critical point of enhancement region is approximately located at \((1.17, 0.125)\). Unlike the reduction region, where \(\tau_r\) changes monotonically with \(\phi\) when \(l\) is fixed, in the enhancement region, \(\tau_r\) first decreases and then increases as \(\phi\) decreases. This behavior indicates that for a given \(l\), there is always a specific value of \(\phi\) that maximizes the enhancement effect, which is plotted as yellow scatters in Figure \ref{diagram}.

\begin{figure}
\centerline{\includegraphics[width=\textwidth]{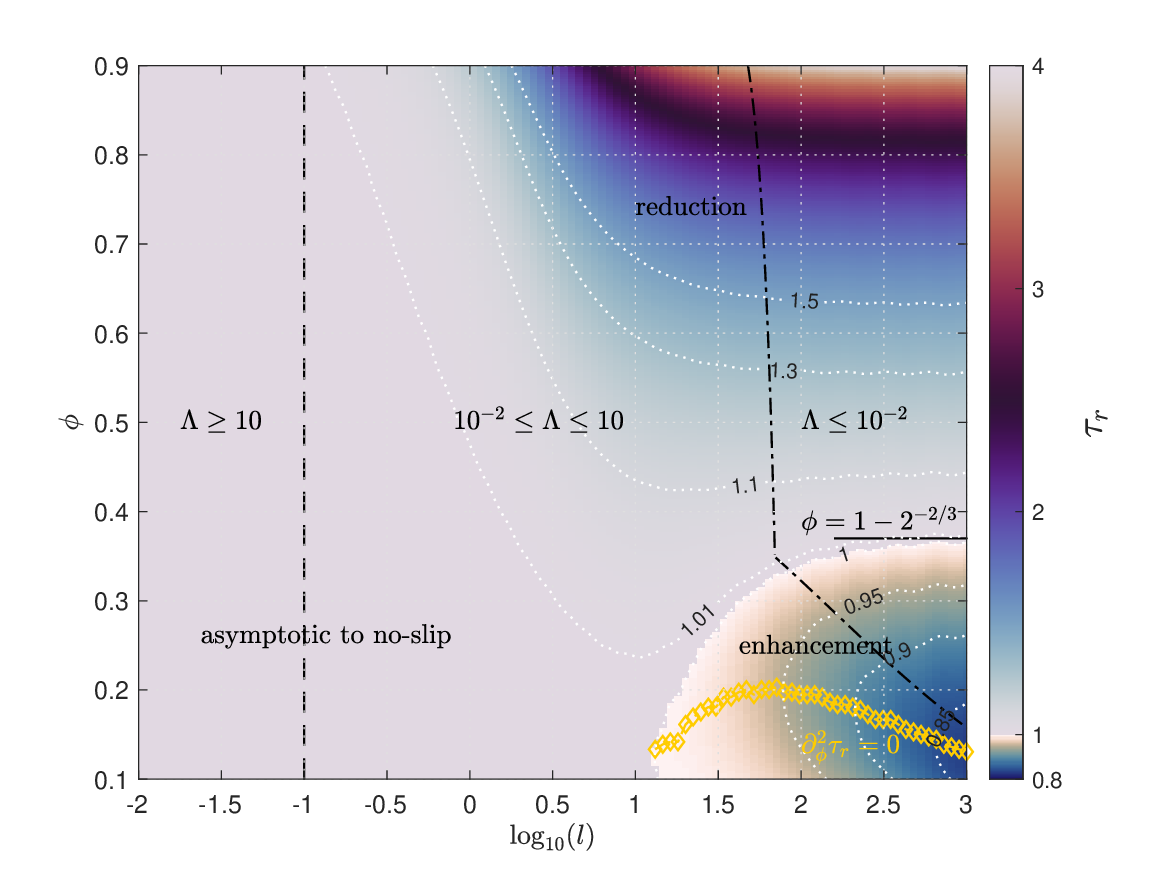}} 
\captionsetup{justification=justified}
    	\caption{Phase diagram of melting time ratio $\tau_r$ of gravity-driven CCM with various combinations of gas-liquid fraction \(\phi\) and period length \(l\) on longitudinal grooves, where white dotted lines are contour lines for various $\tau_r$, orange scatter represents minimum $\tau_r$ at specific $l$, black solid line: $\phi = 1 - 2^{-2/3}$, black dashed line: the asymptotic limit of $l\lessapprox0.1$ for no-slip solutions (\ref{H_tau_no_slip_asymp}), black dot-dash line: the asymptotic limits of (\ref{black_dot_dash}) and (\ref{white_dot_dash}) for solution (\ref{H_lambda_001}).
 }
	\label{diagram}
\end{figure}
Based on this phase diagram, we can further find two regions corresponding to two asymptotes (see details in Appendix \S\ref{asymptotes_H_tau}). 
One is related to asymptotically no-slip solution for $\Lambda\ge10$, as
\begin{equation}
    H(\tau) = \left(1-\frac{3}{4}\tau\right)^{\frac{4}{3}}, \quad l \lessapprox0.1
    \label{H_tau_no_slip_asymp}
\end{equation}
Another asymptotic solution satisfies $\Lambda \lessapprox 0.01$, as
\begin{equation}
     H(\tau;\phi)= \left(1-\frac{3\sqrt{2}}{4}\tau(1-\phi)^{\frac{3}{4}} \right)^{\frac{4}{3}}
     \label{H_lambda_001}
\end{equation}
whose approximate limit conditions should be satisfied as
\begin{subequations}
    \begin{equation}
     100\left(\frac{1-\phi}{4}\right)^{\frac{1}{5}} \lessapprox l
     \label{black_dot_dash}
\end{equation}
\begin{equation}
    l\left[{\frac{4}{3}\sin\theta\frac{\phi^2}{(1-\phi)(1-2\phi)}}\right] \gtrapprox 10
    \label{white_dot_dash}
\end{equation}
\end{subequations}
Therefore, the upper right region enclosed by the two dash-dot lines in Figure \ref{diagram} represents the parameter range where asymptotic solution (\ref{H_lambda_001}) is valid.

\begin{figure}
\captionsetup{justification=justified}
\centerline{\includegraphics[width=0.9 \textwidth]{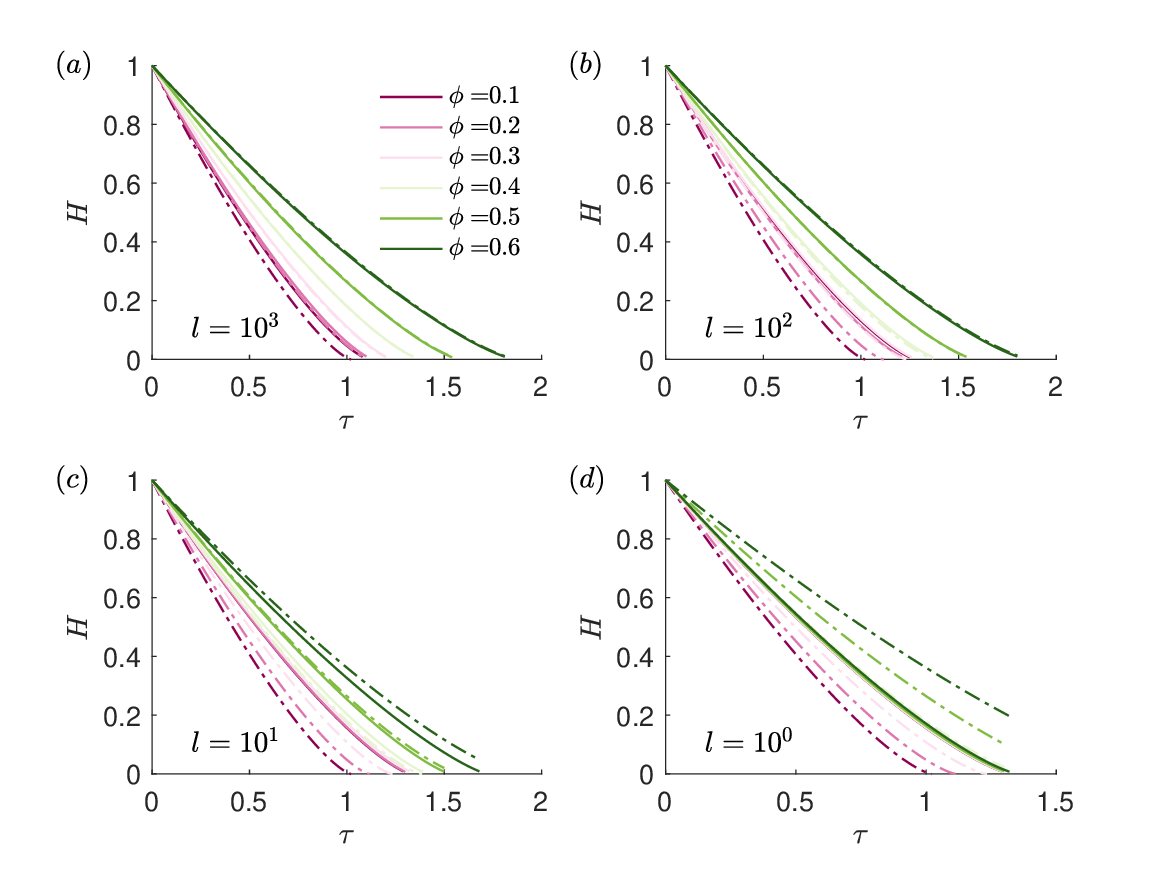}}
    	\caption{Nondimensional height \( H(\tau) \) of solid PCM, comparing numerical results with asymptotic solutions (\ref{H_lambda_001}), at \( l = \) (a) \( 10^3 \), (b) \( 10^2 \), (c) \( 10^1 \), and (d) \( 10^0 \) for \(\phi\) values ranging from 0.1 to 0.6. Solid lines represent numerical results, while dot-dash lines indicate asymptotic solutions.}
	\label{asymptotic_H_tau}
\end{figure}

Comparison of numerical and asymptotic results from $l = 1-10^3$ are plotted in Figure \ref{asymptotic_H_tau}, where solid lines are numerical and dot-dash lines are asymptotic. It shows a good agreement with predictions in phase diagram results by selecting $\phi=\{0.1,0.2,0.3,0.4,0.5,0.6\}$ for cases.

\section{Conclusions}
In this work, we established a theoretical framework to investigate the dynamics of close-contact melting (CCM) on gas-trapped hydrophobic surfaces, with a particular focus on the effects of liquid film confinement and the meniscus at the gas-liquid interface.
Utilizing dual-series and perturbation methods, we obtained numerical results for the velocity slip length \(\lambda\) and temperature slip length \(\lambda_t\) across various aspect ratios \(\Lambda\) and gas fraction $\phi$, representative of the liquid film thickness \(h\) relative to the  period \(l\) of the microstructure. For \(\Lambda \gg 1\), the scaling laws are \(\lambda_{\parallel,f} = \lambda_t = 2\lambda_{\perp,f} = \ln\left(\sec(\phi\pi/2)\right)/\pi\), while for \(\Lambda \ll 1\), they are \(\lambda_{\parallel,f} = \lambda_t = 4\lambda_{\perp,f} = \phi \Lambda/(1-\phi)\), where \(f\), \(\parallel\), and \(\perp\) denote a flat interface, longitudinal grooves, and transverse grooves, respectively. Accounting for the meniscus at the gas-liquid interface, the velocity slip length changes to (\ref{total_lambda_curved_gg_1})  and (\ref{total_lambda_curved_0}), showing a slight decrease for \(\Lambda \gg 1\) and a significant increase for \(\Lambda \ll 1\) compared to the flat interface. Additionally, a modified asymptotic solution for \(\lambda_{\parallel,c}\) (\ref{total_lambda_curved}) is proposed, which is broadly applicable in the range \(\Lambda \gtrapprox 0.2\).

For constant-pressure CCM, the dimensionless film thickness \(h\) remains constant for a specific pressure \(\mathcal{P}\) and both slip lengths, \(\lambda\) and \(\lambda_t\), lead to a slip-affective \(Nu\) that determines whether heat transfer is enhanced (\(Nu > 1\)) or decreased (\(Nu < 1\)). When the gas-liquid interface is flat, we found that transverse grooves surfaces always deteriorate heat transfer regardless of \(\Lambda\) and \(\phi\), while longitudinal grooves exhibit behavior identical to a smooth surface at small \(\phi\) but show deteriorated performance at large \(\phi\). In the presence of a meniscus at the gas-liquid interface, we observed an enhanced melting rate when \(\Lambda \lessapprox 0.1\) and \(\phi < 1-\sqrt[3]{2}/2 \approx 0.37\). It is important to note that \(Nu\) does not vary monotonically with \(\phi\); instead, maximum values of \(Nu\) occur under specific conditions within the range \(\phi < 0.37\) seen Figure \ref{K-map}c.

In gravity-driven CCM, the film thickness \(h\) increases monotonically in time. Slip effects have a minor influence on \(h\)
and become negligible when \(\Lambda \gtrapprox 10\) or when there is a combination of small \(\phi\) and \(\Lambda \sim 10^{-1} - 10\). Conversely, significant slip effects are observed for smaller \(\Lambda\), particularly for \(\Lambda \lessapprox 10^{-2}\). We developed a two-dimensional phase diagram based on 
\((\log_{10} l, \phi)\) to identify regions of enhancement, reduction, or negligible impact of the gas-liquid boundaries (the generator of slip) relative to the no-slip case. Additionally, we derived asymptotic solutions and their limiting conditions.

The results reveal enhanced heat transfer and accelerated melting power in CCM are achievable only when meniscus are present and there are significant confinement effects. The critical conditions depend on the gas-liquid interface fraction being less than 0.37. Therefore, the conditions for utilizing gas-trapped hydrophobic surfaces to enhance CCM melting rates are quite stringent. Future work may explore the potential for heat transfer enhancement using liquid-infused slippery surfaces \citep{Hardt2022} or polymer brush surfaces \citep{Chen2023} to significantly reduce the impact of effective thermal slip while maintaining velocity slip.

\backsection[Acknowledgements]{We thank Fernando Temprano-Coleto for helpful discussions on numerical approaches to the slip length.}

\backsection[Funding]{L.-W. F. acknowledges the 
grant no. 52276088 from National Natural Science Foundation of China. }

\backsection[Declaration of interests]{The authors report no conflict of interest.}

\backsection[Author ORCID]{Nan Hu, https://orcid.org/0000-0002-9377-3904; Li-Wu Fan, https://orcid.org/0000-0001-8845-5058; Howard A. Stone, https://orcid.org/0000-0002-9670-0639.}


\appendix
\setcounter{figure}{0}
\renewcommand{\thefigure}{A\arabic{figure}}

\section{Procedure for solving for the thermal slip length $\lambda_{t,f}$}
\label{Procedure_lambda_t_f}
The general solution \(\tilde{T}\) with condition  (\ref{T_partial_z}) is
\begin{equation}
	\begin{aligned}
		\tilde{T} (y, z) = c_0 \frac{y}{\Lambda} + d_0 + \sum_{n=1}^\infty \left[c_n \cosh \left(2\pi n y\right) + d_n \sinh \left(2\pi n y\right)\right] \cos \left(2\pi n z\right),
	\end{aligned}
\end{equation}
where \(c_n\) and \(d_n\) are constants to be determined. Equation (\ref{long_T_BC_1_imhomo}) results in 
\begin{equation}
	d_0 = -c_0 , \quad d_n = -c_n \coth{(2\pi n \Lambda)} ,
\end{equation}
while equation (\ref{long_T_BC_2_imhomo}) yields 
\begin{equation}
	c_0 + \sum_{n=1}^\infty c_n \alpha_n^T \cos \left(2\pi n z\right) = 0, \quad \phi / 2 < |z| \le 1 / 2
	\label{T_1}
\end{equation}
\begin{equation}
	c_0 + \sum_{n=1}^\infty c_n \beta_n^T\cos \left(2\pi n z\right) = \Lambda, \quad |z| \le \phi / 2,
	\label{T_2}
\end{equation}
where $\alpha_n^T=-1$ and $\beta_n^T=-2 \pi n \Lambda\coth(2 \pi n\Lambda)$.

{Next, we multiply both equations (\ref{T_1}) and (\ref{T_2}) by \(\cos (2 m \pi z)\) for \(m \in [0, N]\), where \(N\) is chosen to numerically truncate the summation. We then integrate the equations after multiplication over the interval \(\phi / 2 < z \leq 1 / 2\) and the interval \(z \leq \phi / 2\), respectively.} Finally, we sum the results to obtain dual-series algebraic equations as:
\begin{subequations}
\begin{equation}
	c_0+\sum_{n=1}^{N} c_n \frac{\beta^{T}_n-\alpha^{T}_n}{\pi n} \sin \left(n \pi \phi \right)=\Lambda\phi \quad \text{for} \quad m=0,
	\label{long_T_2}
\end{equation}
\begin{equation}
	\begin{aligned}
		\sum_{n=1,\neq m}^{N} c_n
		\left[\left(\beta_n^T-\alpha_n^T\right)\frac{m\cos(\pi n\phi)\sin(\pi m\phi)-n\cos(\pi m\phi)\sin(\pi n\phi)}{(m-n)(m+n)\pi} \right] \\
  + c_m
		\left[\frac{\alpha^{T}_m}{2}+\left(\beta^{T}_m-\alpha^{T}_m\right)\left(\frac{\phi}{2}+\frac{\sin (2 \pi m \phi)}{4 \pi m}\right)\right] 
		= \frac{\Lambda\sin (\pi m \phi)}{ \pi m}
		\quad \text{for} \quad m>0.
		\label{long_T_4}
	\end{aligned}
\end{equation}
\end{subequations}
Then we can write the equations in  matrix form as
\begin{equation}
\left[\begin{array}{cccc}
1 & A_{0,1} & \cdots & A_{0, N} \\
0 & A_{1,1} & \cdots & A_{1, N} \\
\vdots & \vdots & \ddots & \vdots \\
0 & A_{N,1} & \cdots & A_{N, N}
\end{array}\right]\left[\begin{array}{c}
c_0 \\
c_1 \\
\vdots \\
c_{N}
\end{array}\right]=\left[\begin{array}{c}
\Lambda \phi \\
\Lambda \sin (\pi \phi)/\pi \\
\vdots \\
\Lambda \sin (\pi N \phi)/\pi N
\end{array}\right]
\end{equation}
where $A_{m,n}$  is defined  for $m=0$, $m=n$ or $m \neq n$:
\begin{equation}
\begin{aligned}
& A_{0,n}=\frac{\beta_n^T-\alpha_n^T}{ \pi n} \sin (n \pi \phi) \\
& \left.A_{n,n}=\frac{\alpha_n^T}{2}+\left(\beta_n^T-\alpha_n^T\right)\left(\frac{\phi}{2}+\frac{\sin (2 \pi n \phi)}{4 \pi n}\right)\right) \\
& A_{m,n}=\frac{m \cos (\pi n \phi) \sin (\pi m \phi)-n \cos (\pi m \phi) \sin (\pi n \phi)}{(m-n)(m+n) \pi}.
\end{aligned}
\end{equation}

\section{Thermal slip length $\lambda_{t,c}$ on curved gas-liquid interface}
\label{Procedure_lambda_t_c}
Substituting the perturbation expansion for  temperature $T=T^{(0)}+\epsilon T^{(1)}+O\left(\epsilon^2\right)$ into (\ref{T_curved_BC}) we obtain a series of equations at different orders of $\epsilon$.
At order of $O(\epsilon^0)$, we have 
\begin{equation}
	\partial_{y y} T^{(0)} + \partial_{zz} T^{(0)}=0,
\end{equation}
with the boundary conditions as
\begin{equation}
	\begin{aligned}
		& T^{(0)}(\Lambda, z)=0\\
		& \left\{\begin{array}{ll}
			T^{(0)}(0, z)=0, & \phi/2<|z| \leqslant 1 / 2 \\
			\partial_y T^{(0)}(0, z)=0, & |z| \leqslant \phi/2,
		\end{array} \right.
	\end{aligned}
\end{equation}
which is exactly the case of flat interface and has been  solved in \S \ref{sec_lambda_t_f} and Appendix \S \ref{Procedure_lambda_t_f}, namely $T^{(0)}=T_p+T_s$.

At $O(\epsilon^1)$, the problem is formulated as 
\begin{equation}
	\partial_{y y} T^{(1)}+\partial_{zz} T^{(1)}=0,
\end{equation}
with the boundary conditions
\begin{equation}
	T^{(1)}(\Lambda, z)=0
	\label{T^{1} upper BC}
\end{equation}
\begin{equation}
	\left\{\begin{array}{ll}
		T^{(1)}(0, z)=0, & \phi/2<|z| \leqslant 1 / 2 \\
		\partial_y T^{(1)}(0, z)=\eta \partial_{y y} T^{(0)}(0, z)-\eta^{\prime} \partial_z T^{(0)}(0, z), & |z| \leqslant \phi/2  .
	\end{array} \right.
\label{T^{1} bottom BC}
\end{equation}
The general solution of the first-order temperature ${T}^{(1)}=\tilde{T}^{(1)}/\Lambda$, where $\tilde{T}^{(1)}$ has the same form as (\ref{T_general_form}), i.e.
\begin{equation}
		\tilde{T}^{(1)} (y, z)= c_0^{(1)} \frac{y}{\Lambda} + d_0^{(1)} + \sum_{n=1}^\infty \left[c_n^{(1)} \cosh \left(2\pi n y\right) + d_n^{(1)} \sinh \left(2\pi n y\right)\right] \cos \left(2\pi n z\right).
\end{equation}

Applying boundary condition (\ref{T^{1} upper BC}) give
\begin{equation}
	{T}^{(1)}=c_0^{(1)}(y-\Lambda)+\sum_{n=1}^{\infty} c_n^{(1)} \frac{\sinh \left[2\pi n(\Lambda-y)\right]}{\sinh \left(2\pi n \Lambda \right)} \cos \left(2\pi n z\right).
	\label{T^{1}}
\end{equation}

Then, applying conditions (\ref{T^{1} bottom BC}) in (\ref{T^{1}}) yields
\begin{equation}
	\begin{gathered}
		c_0^{(1)}+\sum_{n=1}^{\infty} c_n^{(1)} \alpha^{T}_n
		\cos \left(2\pi n z\right)=0, \quad \phi /2 <|z| \le 1/2, \\
	c_0^{(1)}+\sum_{n=1}^{\infty} c_n^{(1)} \beta^{T}_n \cos \left(2\pi n z\right)=
		-\Lambda\partial_z\left(\eta \partial_z \tilde{T}^{(0)}(0, z)\right), \quad |z| \le \phi/2 ,
	\end{gathered}
\end{equation}
where
\begin{equation}
	-\Lambda \partial_z\left(\eta \partial_z \tilde{T}^{(0)}(0, z)\right) = 
	\sum_{n=1}^{\infty} \Lambda  c_n 
	 \left[2\pi n \sin \left(2\pi n z\right)\eta^{\prime}+(2\pi n)^2 \cos \left(2\pi n z\right)\eta\right].
\end{equation}
Similarly, dual series equations can be obtained by multiplying these equations by $\cos\left(2\pi m z\right)$, integrating over the corresponding domain and summing them. The dual series equations also can be solved numerically by the same procedure as in  Appendix \S \ref{Procedure_lambda_t_f}. Nevertheless, we can immediately find
\begin{equation}
		c_0^{(1)}=0,
\end{equation}
because every term equals to 0 for $m=0$.
Now, we can notice that mean heat flux $\left< q^{\prime\prime}\right>_c$ across $y=\Lambda$ is 
\begin{equation}
    \left< q^{\prime\prime}\right>_c = \int_{-1/2}^{1/2}-\partial_yT(\Lambda,z)dz= \frac{1}{\Lambda}-\frac{c_0}{\Lambda^2}-\epsilon\frac{c_0^{(1)}}{\Lambda^2}=\frac{1}{\Lambda}-\frac{c_0}{\Lambda^2},
\end{equation}
which reveals that the presence of the meniscus has no influence on the average heat flux, i.e., $\lambda_{t,c}=\lambda_{t,f}$.

\section{Procedure for solving for the longitudinal slip length $\lambda_{\parallel,f}$}
\label{Procedure_lambda_perp_f}
The general solution of $\tilde{u}$ along the parallel grooves is
\begin{equation}
	\begin{aligned}
		\tilde{u} (y,z) =  r_0 +q_{0}\frac{y}{\Lambda}+\sum_{n=1}^\infty \left[r_n\cosh \left(2\pi n y\right)+q_n \sinh \left(2\pi n y\right)\right] \cos \left(2\pi n z\right),
		\label{deviation solution of w}
	\end{aligned}
\end{equation}
where $r_n$ and $q_n$ are constants to be determined.
Substitution of (\ref{long_BC_1}) results in 
\begin{equation}
	q_0=-r_0, \quad q_n=-r_n \coth \left(2\pi n\Lambda\right).
\end{equation}
Then (\ref{long_BC_2}) results in 
\begin{subequations}
    \begin{equation}
	r_0+\sum_{n=1}^{\infty} r_n\alpha^{\|}_n \cos \left(2\pi n z\right)=0, \quad  \phi /2 <|z| \le 1/2
	\label{para_eq_1}
\end{equation}
\begin{equation}
	r_0+\sum^{\infty} r_n\beta^{\parallel}_n \cos \left(2\pi n z\right)=\Lambda,\quad |z| \le \phi /2
	\label{para_eq_2}
\end{equation}
\end{subequations}
The coefficients of $\alpha^{\parallel}_n$ and $\beta^{\parallel}_n$ are
\begin{equation}
	\begin{aligned}
		\alpha^{\|}_n & =1  \\
		\beta^{\|}_n & =2\pi n\Lambda\coth(2\pi n\Lambda).
	\end{aligned}
\end{equation}
Following what are now standard steps, we also can obtain the dual-series algebraic equations as 
\begin{subequations}
\begin{equation}
	r_0+\sum_{n=1}^{\infty} r_n \frac{\beta^{\parallel}_n-\alpha^{\parallel}_n}{\pi n} \sin \left(n \pi \phi \right)=\Lambda\phi \quad \text{ for } m=0,
	\label{para_1}
\end{equation}
\begin{equation}
	\begin{aligned}
		\sum_{n=1,\neq m}^{\infty} r_n \left[\left(\beta_n^{\parallel}-\alpha_n^{\parallel}\right) \frac{m\cos(\pi n\phi)\sin(\pi m\phi)-n\cos(\pi m\phi)\sin(\pi n\phi)}{(m-n)(m+n)\pi}\right] \\
  + r_m
		\left[\frac{\alpha^{\parallel}_m}{2}+\left(\beta^{\parallel}_m-\alpha^{\parallel}_m\right)\left(\frac{\phi}{2}+\frac{\sin (2 \pi m \phi)}{4 \pi m}\right)\right] = \frac{\Lambda\sin (\pi m \phi)}{ \pi m}\quad \text{ for } m>0.
		\label{para_2}
	\end{aligned}
\end{equation}
\end{subequations}
(\ref{para_1} - \ref{para_2}) can be numerically solved for $r_0$ and $r_n$ for $m \in[0, N]$ and $n \in[0, N]$. 

\section{Procedure for solving $\lambda_{\parallel.c}$}
\label{Procedure_lambda_perp_f_c}
Substituting the perturbation velocity $u=u^{(0)}+\epsilon u^{(1)}+O\left(\epsilon^2\right)$ into (\ref{velocity_curve_BC}) we obtain a series of equations at different orders of $\epsilon$.
At order of $O(\epsilon^0)$, we have
\begin{equation}
	\partial_{zz} u^{(0)}+\partial_{y y} u^{(0)}=-\partial_xP,
\end{equation}
\begin{equation}
	\begin{aligned}
		& u^{(0)}(\Lambda, z)=0\\
		& \left\{\begin{array}{ll}
			u^{(0)}(0, z)=0, & \phi/2<|z| \leqslant 1 / 2 \\
			\partial_y u^{(0)}(0, z)=0, & |z| \leqslant \phi/2
		\end{array} \right. ,
	\end{aligned}
\end{equation}
which is the case of the flat meniscus and has been solved as shown in Appendix \S \ref{Procedure_lambda_perp_f}. Hence $r_n^{(0)}=r_n$ and
\begin{equation}
	\lambda_{\parallel,c}^{(0)} = \lambda_{\parallel,f}=\frac{\Lambda r_0}{\Lambda - r_0},
\end{equation}
where \(r_0\) is the coefficient determined by numerically solving equations (\ref{para_1} - \ref{para_2}).

At  $O(\epsilon^1)$, the equations are 
\begin{equation}
	\partial_{zz} u^{(1)}+\partial_{y y} u^{(1)}=0,
\end{equation}
\begin{equation}
	u^{(1)}(\Lambda, z)=0
	\label{u^{1} upper BC}
\end{equation}
\begin{equation}
	\left\{\begin{array}{ll}
		u^{(1)}(0, z)=0, & \phi/2<|z| \leqslant 1 / 2 \\
		\partial_y u^{(1)}(0, z)=\eta \partial_{y y} u^{(0)}(0, z)-\eta^{\prime} \partial_z u^{(0)}(0, z), & |z| \leqslant \phi/2
	\end{array} \right.
\label{u^{1} bottom BC}
\end{equation}
The general solution of first-order velocity ${u}^{(1)}=-\partial_xP\Lambda \tilde{u}^{(1)}/2$, where $\tilde{u}^{(1)}$ has the same form as (\ref{deviation solution of w}), i.e.,
\begin{equation}
	\tilde{u}^{(1)}=r_0^{(1)} +q_0^{(1)}\frac{y}{\Lambda}+\sum_{n=1}^{\infty}\left[r_n^{(1)} \cosh \left(2\pi n y\right)+q_n^{(1)} \sinh \left(2\pi n y\right)\right] \cos \left(2\pi n z\right)
\end{equation}
Applying the boundary condition of (\ref{u^{1} upper BC}) leads to
\begin{equation}
	\tilde{u}^{(1)}(y,z)=r_0^{(1)}(1-\frac{y}{\Lambda})+\sum_{n=1}^{\infty} r_n^{(1)} \frac{\sinh \left[2\pi n(\Lambda-y)\right]}{\sinh \left(2\pi n \Lambda \right)} \cos \left(2\pi n z\right).
	\label{u^{1}}
\end{equation}
Then applying condition (\ref{u^{1} bottom BC}) in (\ref{u^{1}}) yields
\begin{equation}
	\begin{gathered}
		r_0^{(1)}+\sum_{n=1}^{\infty} r_n^{(1)} \alpha^{\|}_n
		\cos \left(2\pi n z\right)=0, \quad \phi /2 <|z| \le 1/2, \\
	r_0^{(1)}+\sum^{\infty} r_n^{(1)} \beta^{\parallel}_n \cos \left(2\pi n z\right)=
		\Lambda\partial_z\left(\eta \partial_z \tilde{u}^{(0)}(0, z)\right)+2\eta, \quad |z| \le \phi/2 ,
	\end{gathered}
\end{equation}
where
\begin{equation}
	\partial_z\left(\eta \partial_z \tilde{u}^{(0)}(0, z)\right) = 
	\sum_{n=1}^{\infty} r_n 
	 \left[-2\pi n \sin \left(2\pi n z\right)\eta^{\prime}-(2\pi n)^2 \cos \left(2\pi n z\right)\eta\right] .
\end{equation}
Similarly, dual series equations can be obtained by multiplying $\cos\left(2\pi m z\right)$ and integrating. The dual series equations also can be solved numerically using the same procedure as in Sec. \S \ref{Procedure_lambda_t_f}:
\begin{subequations}
\begin{equation}
	r_0^{(1)}+\sum_{n=1}^{\infty} r_n^{(1)} \frac{\beta^{\parallel}_n-\alpha^{\parallel}_n}{\pi n} \sin \left(n \pi \phi \right)=\frac{4\phi^3}{3} \quad \text{for} \quad m=0,
	\label{para_2_curve}
\end{equation}
\begin{equation}
	\begin{aligned}
		&\sum_{n=1,\neq m}^{\infty} r_n^{(1)}
		\left[\left(\beta_n^{\parallel}-\alpha_n^{\parallel}\right) \frac{1}{2 \pi}\left(\frac{\sin (\pi(m+n) \phi)}{m+n}+\frac{\sin (\pi(m-n) \phi)}{m-n}\right)\right] \\
  &+ r_m^{(1)}
		\left[\frac{\alpha^{\parallel}_m}{2}+\left(\beta^{\parallel}_m-\alpha^{\parallel}_m\right)\left(\frac{\phi}{2}+\frac{\sin (2 \pi m \phi)}{4 \pi m}\right)\right] 
		= \mathcal{M}\left(n,m,\Lambda,\phi\right)
		\quad  \text{for} \quad m>0,
		\label{para_4_curve}
	\end{aligned}
\end{equation}
\end{subequations}
where the function $\mathcal{M}\left(n,m,\Lambda,\phi\right)$ is
\begin{subequations}
\begin{equation}
	\begin{aligned}
		\mathcal{M}\left(n,m,\Lambda,\phi\right) &=  \,  \frac{4\sin(\pi \phi m) - 4\pi \phi m \cos(\pi \phi m)}{ \pi^3 m^3} +\sum_{n=1,\neq m}^{\infty}\frac{4r_n \Lambda m n}{(m-n)^3(m+n)^3 \pi}\\
		&\left\{m\cos(m \pi \phi)\left[2 n\left(m^2-n^2\right) \pi \phi \cos (n \pi \phi)+\left(m^2+3 n^2\right) \sin (n \pi \phi)\right]+\right. \\
		& \left.\sin (m \pi \phi)\left[-n\left(3 m^2+n^2\right) \cos (n \pi \phi)+\left(m^4-n^4\right) \pi \phi \sin (n \pi \phi)\right]\right\}+\\
        & r_m\Lambda \left(
		\frac{3 \sin(2 m \pi \phi)-8 m^3 \pi^3 \phi^3-6 m \pi \phi \cos (2 m \pi \phi)}{12 m \pi}
		\right)
	\end{aligned}
\end{equation}
\end{subequations}

\section{Slip length $\lambda_{\perp,f}$ on transverse grooves with flat liquid-gas interface}\label{lambda_perp}

The general solution of deviation component of the streamfuction $\tilde{\Psi}$ between two parallel plate with transverse grooves, is well known as \citep{RN1091,RN1046}
\begin{equation}
	\begin{aligned}
		\tilde{\Psi}= & C_0 y+\frac{D_0 y^2}{2\Lambda}+\sum_{n=1}^{\infty}\left\{C_n\left[\cosh \left(2\pi n y\right)-\operatorname{coth}\left(2\pi n\right) y \sinh \left(2\pi n y\right)\right]\right. \\
		& \left.+D_n\left[\sinh \left(2\pi n y\right)-\tanh \left(2\pi n\right) y \cosh \left(2\pi n y\right)\right]\right\} \cos \left(2\pi n z\right) ,
	\end{aligned}
\end{equation}
where $x$ and $y$ are,  respectively, the coordinates along width and height in the cross-section of flow, and $C_0$, $D_0$, $C_n$ and $D_n$ are unknown coefficients determined by specific boundary conditions;$2\pi n $ represents the wave number. The boundary conditions for grooves on a single channel wall are
\begin{equation}
	\partial_y \tilde{\Psi}|_{y=\Lambda}=0 
	\label{tran_BC_1}
\end{equation} 
\begin{equation}
 \begin{cases}
 	\partial_y \tilde{\Psi}(x,0)=0, & \phi /2 <|z| \le 1/2 \\ 1+\partial_{yy} \tilde{\Psi}(x,0)=0, & |z|\le \phi /2
 \end{cases}
 	\label{tran_BC_2}
\end{equation} 
Equation (\ref{tran_BC_1}) leads to 
\begin{equation}
	C_0 = -D_0,\quad C_n=D_n \gamma,
\end{equation}
where 
\begin{equation}
	\gamma=-\frac{
		2\pi n \cosh \left(2\pi n\Lambda\right)-
		\tanh \left(2\pi n\right)\cosh\left(2\pi n\Lambda\right)-
		2\pi n\Lambda \tanh \left(2\pi n\right) \sinh \left(2\pi n\Lambda\right)}
	{2\pi n \sinh \left(2\pi n\Lambda\right)-\coth \left(2\pi n\right)\sinh\left(2\pi n\Lambda\right)-2\pi n\Lambda \operatorname{coth}\left(2\pi n\right) \cosh \left(2\pi n\Lambda\right)}.
\end{equation}
\\
Also, (\ref{tran_BC_2}) leads to 
\begin{equation}
	C_0+\sum_{n=1}^{\infty} C_n \alpha^{\perp}_{n} \cos \left(2\pi n z\right)=0
	\label{tran_eq_1}
\end{equation}
for $\phi /2 <|z| \le 1/2$ and
\begin{equation}
	C_0+\sum_{n=1}^{\infty} C_n \beta^{\perp}_{n} \cos \left(2\pi n z\right)=\Lambda
		\label{tran_eq_2}
\end{equation}
for $|z| \le \phi /2$. The coeffienct $\alpha^{\perp}_{n}$ and $\beta^{\perp}_{n}$ are 
\begin{equation}
	\begin{aligned}
		\alpha^{\perp}_{n}= 
	\left[	2 n \pi-\tanh\left(2 n \pi\right)\right]/\gamma
	\end{aligned}
\end{equation}
and
\begin{equation}
	\begin{aligned}
		\beta^{\perp}_{n}= 
		\left[4n\pi\coth\left(2n\pi\right)-4n^2\pi^2\right]\Lambda .
	\end{aligned}
\end{equation}

Finally, 
\begin{equation}
	 \lambda_{\perp} = \frac{\Lambda C_0}{\Lambda-C_0}.
\end{equation}

Similar to the approach in the last section, values of $C_0$ and $C_n$ can be numerically solved by replacing $\alpha^{\|}$ and $\beta^{\|}$ with $\alpha^{\perp}$ and $\beta^{\perp}$  respectively.

\section{Numerical approach for solving solid height $H(\tau)$ and film thickness $h(\tau)$ of gravity-driven CCM}
\label{solving_H_tau}
Recalling the equations (\ref{H-h}) and (\ref{H-tau}) with $\mathscr{c}=1$,
we can rewrite them by defining the functions $f(\Lambda)$ and $g(\Lambda)$ respectively, as
\begin{subequations}
    \begin{equation}
    \Lambda^4\frac{1+4\lambda/\Lambda}{1+\lambda/\Lambda}\left(1+\frac{\lambda_{t}}{\Lambda}\right) \equiv f(\Lambda)=\frac{1}{Hl^4}
    \label{f_Lambda}
\end{equation}
\begin{equation}
	\frac{dH}{d\tau}=-\frac{1}{l(\Lambda+\lambda_{t})} \equiv -l^{-1} g(\Lambda)
	\label{g_Lambda}
\end{equation}
\end{subequations}
Using the (\ref{f_Lambda}) and (\ref{g_Lambda}) and the initial condition \(H(\tau=0)=1\), we can numerically compute \(H\) through the following discrete iterative equations
\begin{subequations}
    \begin{equation}
    f(\Lambda^i)=l^{-4}\frac{1}{H^i}
    \label{f_Lambda_num}
\end{equation}
\begin{equation}
	H^{i+1}= -l^{-1} g(\Lambda^{i}) \tau_\delta +H^{i}
 \label{g_Lambda_num}
\end{equation}
\end{subequations}
where the superscript \(i\) represents the time step, and \(\tau_\delta \) denotes the time increment. This implies that \(\Lambda^{i}\) can be determined using \(H^{i}\) at the current time step \(i\) via equation (\ref{f_Lambda_num}). Subsequently, \(H^{i+1}\) can be calculated by substituting \(\Lambda^{i}\) into equation (\ref{g_Lambda_num}). 
\begin{figure}
\captionsetup{justification=justified}
\centerline{\includegraphics[width=0.9\textwidth]{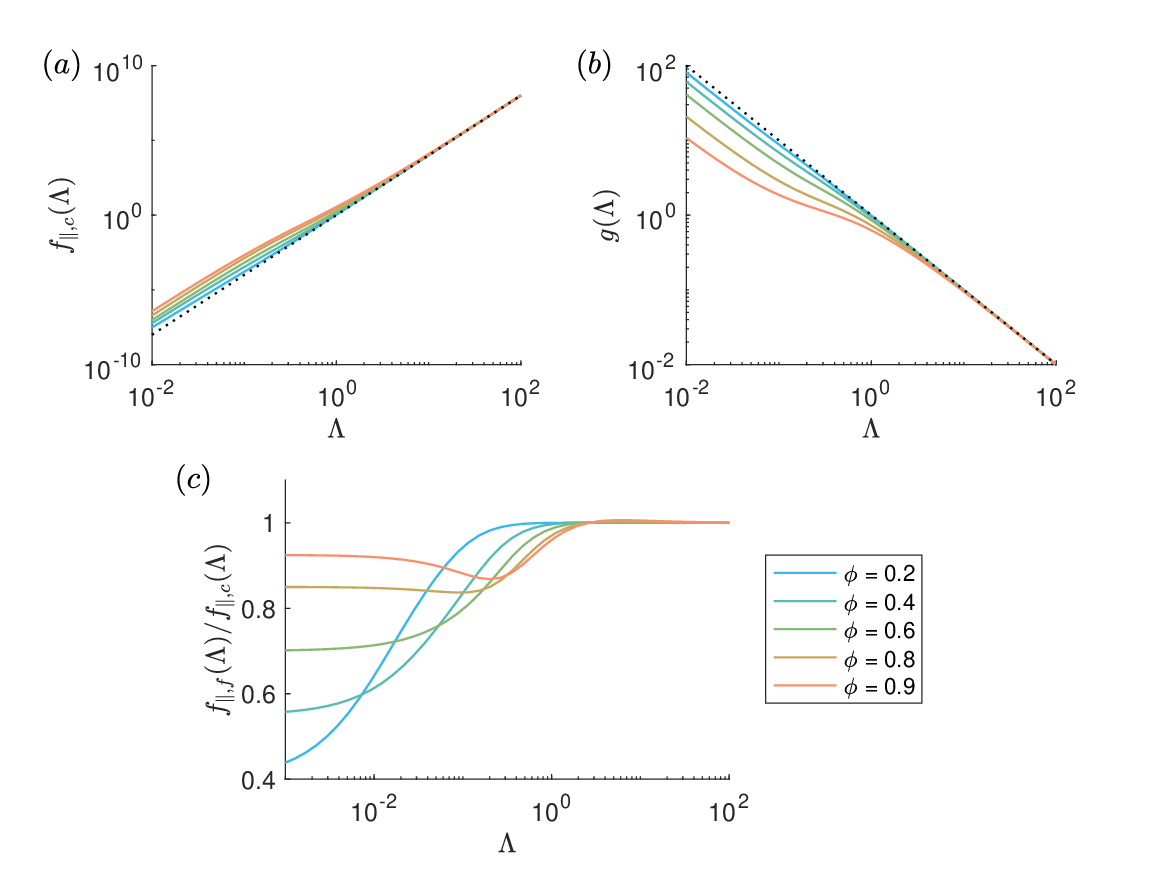}} 
	\caption{ Functions of (a) \(f_{\parallel,c}(\Lambda)\) and (b) \(g(\Lambda)\) for the meniscus interface on longitudinal grooves for iterative numerical approaches, and (c) the ratio \(f_{\parallel,f}(\Lambda)/f_{\parallel,c}(\Lambda)\) of flat to curved interface. Dot line represents no-slip.
 }
	\label{f_g_Lambda}
\end{figure}
However, since the functions \(f(\Lambda)\) and \(g(\Lambda)\) do not have analytical expressions, it is necessary to precompute a sufficient number of discrete values for these functions for a specific \(\phi\). During the iteration process, interpolation is used to obtain \(f(\Lambda)\) and \(g(\Lambda)\) for any given \(\Lambda\) based on discrete results. The functions \(f_{\parallel,c}(\Lambda)\) and \(g(\Lambda)\) for longitudinal grooves, considering the meniscus interface, are plotted in Figures \ref{f_g_Lambda}a and b, respectively. Since the \(f_{\parallel,f}(\Lambda)\) for flat interfaces is very similar to that for meniscus interfaces, Figure \ref{f_g_Lambda}c presents the ratio of \(f_{\parallel,f}(\Lambda)\) for flat interfaces to \(f_{\parallel,c}(\Lambda)\) for meniscus interfaces.
\section{Asymptotic solutions of $H(\tau)$ and corresponding conditions for gravity-driven mode}\label{asymptotes_H_tau}
\begin{figure}
\captionsetup{justification=justified}
    \centering
\includegraphics[width=\linewidth]{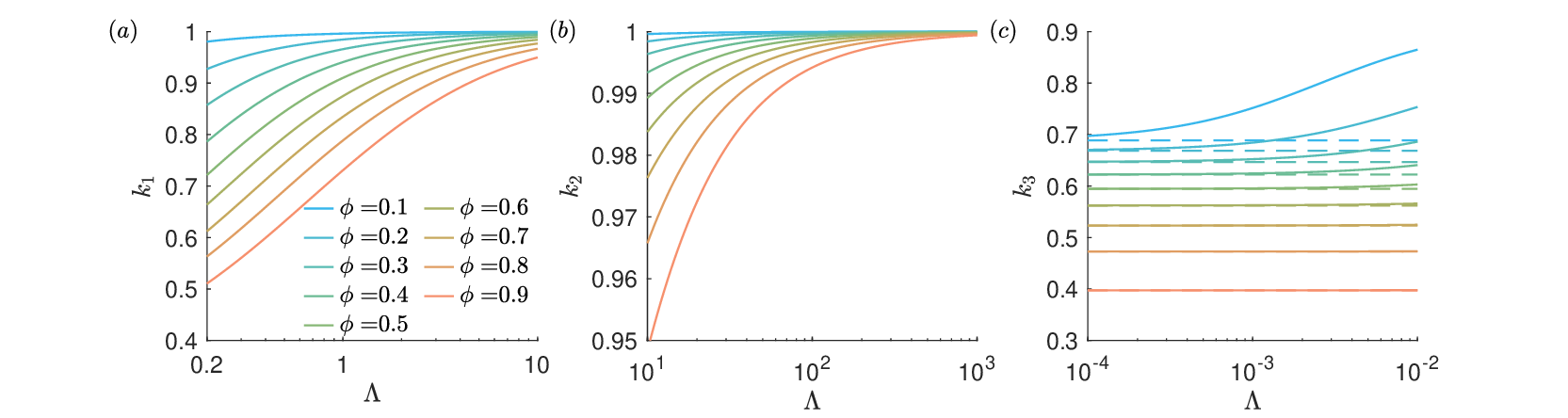}1
\caption{Variations of coefficients (a)$k_1$, (b)$k_2$ and (c)$k_3$ versus aspect ratio $\Lambda$}
    \label{k1_k2_k3}
\end{figure}
With the given asymptotic formulae of slip lengths in Table \ref{asymptotic}, we can derive the asymptotic solutions and their conditions for $H-\tau$.
Considering the range of $\Lambda \gtrapprox 0.2$, substitution of (\ref{total_lambda_curved}) into (\ref{H_Lambda}) yields 
\begin{equation}
    h =  k_{1}H^{-\frac{1}{4}}, \quad \Lambda \gtrapprox 0.2
\end{equation}
where $k_{1}$ is given as
\begin{equation}
\begin{split}
    k_{1} = &\left(\frac{1 + \frac{1}{\Lambda\pi} \ln \left(\sec \left(\frac{\phi \pi}{2}\right)\right) + \frac{\epsilon}{\Lambda} \left(-\phi^3 \mathcal{F}(\phi) + \frac{4}{\Lambda} \phi^4 \mathcal{G}(\phi)\right) \left(1 + \frac{1}{4\Lambda\pi} \ln \left(\sec \left(\frac{\phi \pi}{2}\right)\right)\right)^2}{1 + \frac{4}{\Lambda\pi} \ln \left(\sec \left(\frac{\phi \pi}{2}\right)\right) + \frac{4\epsilon}{\Lambda} \left(-\phi^3 \mathcal{F}(\phi) + \frac{4}{\Lambda} \phi^4 \mathcal{G}(\phi)\right) \left(1 + \frac{1}{4\Lambda\pi} \ln \left(\sec \left(\frac{\phi \pi}{2}\right)\right)\right)^2} \right)^\frac{1}{4} \\
    &\left(1 + \frac{4}{\Lambda\pi} \ln \left(\sec \left(\frac{\phi \pi}{2}\right)\right)\right)^{-\frac{1}{4}}
    \label{Lambda>=0.2}
\end{split}
\end{equation}
and is plotted in Figure \ref{k1_k2_k3}a. It demonstrated that only for \(\Lambda = 1-10\) and small \(\phi\), \(k_{1}\) can be considered as approximately equal to 1, which is consistent with the results in Figures \ref{H_h_t}c and \ref{H_h_t}d. 
Especially, when $\Lambda \gtrapprox 10$ , we can further simplify equation (\ref{Lambda>=0.2}) as follows:
\begin{equation}
    h = \left({1+\frac{4}{\Lambda\pi}\ln(\sec\frac{\phi\pi}{2})}\right)^{-\frac{1}{4}}H^{-\frac{1}{4}} \equiv  k_{2}H^{-\frac{1}{4}},\quad \Lambda \gtrapprox 10
\end{equation}
By numerically analyzing the order of \(k_{\Lambda \ge 10}\), it is found to have a magnitude of \(\sim 1\) within the range \(\Lambda = 10 - 10^{3}\), as depicted in Figure \ref{k1_k2_k3}b, although with a smaller deviation for large \(\phi\). Therefore, it is reasonable to approximate the above equation as
\begin{equation}
h \approx  H^{-\frac{1}{4}} 
\end{equation}
which is exactly equivalent to the formulas (\ref{H_tau_no_slip}) and (\ref{h_tau_no_slip}) for the no-slip condition. To satisfy the condition \(\Lambda \gtrapprox 10\), it follows that
\begin{equation}
l \lessapprox 0.1
\end{equation}
due to the necessary condition \(h(\tau=0)/l \gtrapprox 10\).

As for the range of $\Lambda \lessapprox 0.01$, we can substitute (\ref{total_lambda_curved_0}) into (\ref{f_Lambda}), yielding
\begin{equation}
     h = 
    \left[1+\frac{1}{\frac{1}{3}+\frac{4}{3}\frac{\phi}{1-\phi}+\frac{8\sin\theta}{9\Lambda}\frac{\phi^2}{(1-\phi)^2}}\right]^{\frac{1}{4}}
     \left(\frac{1-\phi}{4}\right)^{\frac{1}{4}}H^{-\frac{1}{4}} \equiv k_{3}H^{-\frac{1}{4}}, \quad    \Lambda \lessapprox 0.01
\end{equation}
By plotting $k_{3}$ along with $\Lambda$ in Figure \ref{k1_k2_k3}c, it demonstrates that $k_{3}$ remains constant for $\phi = 0.5 - 0.9$ while deviates greater for smaller $\phi \le 0.4$. We can  easily obtain the asymptotic profile for $\phi = 0.5 - 0.9$ is $(1-\phi)^{0.25}/4^{0.25}$ as plotted as a dashed line in Figure \ref{diagram}d. Therefore, the maximum film thickness at the end $h_{end}$ is
\begin{equation}
    h_{end} = \left( \frac{1-\phi}{4}\right)^{\frac{1}{5}}
\end{equation}
by letting $H = h_{end}$. Then one limit condition for $h_{end}/l \lessapprox 0.01$ is
\begin{equation}
     1 \lessapprox \left( \frac{l}{100}\right)^5  \frac{4}{1-\phi}
\end{equation}
which is plotted as black dot-dash lines in Figure \ref{diagram}a.
Another limit condition lies in 
\begin{equation}
    \frac{1}{3}+\frac{4}{3}\frac{\phi}{1-\phi}+\frac{8\sin\theta}{9\Lambda}\frac{\phi^2}{(1-\phi)^2} \gg 1
\end{equation}
which can be derived into
\begin{equation}
    \log_{10}{l} \gg \log_{10}{h} -\log_{10}\left[{\frac{4}{3}\sin\theta\frac{\phi^2}{(1-\phi)(1-2\phi)}}\right]
\end{equation}
By adopting $h=h_{max}=10$ from no-slip condition, this condition is plotted as white dot-dash lines in Figure \ref{diagram}a. Therefore, the upper right region enclosed by the two dash-dot lines represents the parameter range where the following asymptotic solution is valid.
\begin{equation}
     H(\tau;\phi)= \left(1-\frac{3\sqrt{2}}{4}\tau(1-\phi)^{\frac{3}{4}} \right)^{\frac{4}{3}}, \quad \Lambda \lessapprox 0.01
\end{equation}
\newpage
\bibliographystyle{jfm}
\bibliography{jfm}

\end{document}